\PassOptionsToPackage{table}{xcolor}
\documentclass{article}

 \usepackage[dandb, final]{neurips_2025}

\usepackage{tikz}
\usetikzlibrary{shapes.geometric} 
\usetikzlibrary{positioning}     
\usetikzlibrary{arrows}  

\tikzset{
  data/.style={rectangle, rounded corners, draw, fill=blue!10, text width=6em, minimum height=2em, align=center},
  process/.style={rectangle, draw, fill=orange!10, text width=8em, minimum height=2em, align=center},
  decision/.style={
      trapezium, 
      trapezium left angle=70, 
      trapezium right angle=110, 
      draw, 
      fill=green!10, 
      text width=7em, 
      minimum height=2em, 
      align=center
  },
  arrow/.style={->, >=stealth, thick}   
}

\usepackage[utf8]{inputenc} 
\usepackage[T1]{fontenc}    
\usepackage{hyperref}       
\usepackage{url}            
\usepackage{booktabs}       
\usepackage{amsfonts}       
\usepackage{nicefrac}       
\usepackage{microtype}      
\usepackage[table]{xcolor} 

\usepackage{multirow}
\usepackage{amsmath} 
\usepackage{booktabs}        
\usepackage{threeparttable}  
\usepackage[table]{xcolor}
\usepackage{multirow}
\usepackage{tabularx}
\usepackage{graphicx}
\usepackage{wrapfig}
\usepackage{subcaption}
\usepackage[most]{tcolorbox}
\usepackage{adjustbox}

\newcommand{\cmark}{\textcolor{green!60!black}{\ding{51}}}  
\newcommand{\xmark}{\textcolor{red}{\ding{55}}}             
\usepackage{pifont}  

\title{CPRet: A Dataset, Benchmark, and Model for Retrieval in Competitive Programming}

\author{
Han Deng$^{1,2,3,5}$, 
Yuan Meng$^{3,*}$, 
Shixiang Tang$^2$, 
Wanli Ouyang $^{1,2,5}$, 
Xinzhu Ma$^{2,4,}$\thanks{Corresponding author.} \\
$^1$ The Chinese University of Hong Kong \quad
$^2$ Shanghai Artificial Intelligence Laboratory \\
$^3$ Tsinghua University \quad
$^4$ Beihang University  \quad
$^5$ Shenzhen Loop Area Institute
}

\begin{document}

\maketitle

\begin{abstract}
Competive programming benchmarks are widely used in scenarios such as programming contests and large language model assessments. 
However, the growing presence of duplicate or highly similar problems raises concerns not only about competition fairness, but also about the validity of competitive programming as a benchmark for model evaluation. 
In this paper, we propose a new problem—similar question retrieval—to tackle the problem.
Due to the lack of both data and models, solving this problem is challenging.
To this end, we introduce \textbf{CPRet}, a retrieval-oriented benchmark suite for competitive programming, covering four retrieval tasks: two code-centric ({\it i.e.}, {\it Text-to-Code}, {\it Code-to-Code}) and two newly proposed problem-centric tasks ({\it i.e.}, {\it Problem-to-Duplicate}, {\it Simplified-to-Full})—built from a combination of automatically crawled problem–solution data and manually curated annotations. Our contribution includes both high-quality training data and temporally separated test sets for reliable evaluation. Besides, we further develop two task-specialized retrievers based on this dataset: CPRetriever-Code, trained with a novel Group-InfoNCE loss for problem–code alignment, and CPRetriever-Prob, fine-tuned for indentifying problem-level similarity. Both models achieve strong results and are open-sourced for local use. Finally, we analyze LiveCodeBench and find that high-similarity problems inflate model pass rates and reduce differentiation, underscoring the need for similarity-aware evaluation in future benchmarks.
\end{abstract}

\begin{center}
\textbf{Github:} \href{https://github.com/coldchair/CPRet}{https://github.com/coldchair/CPRet}

\textbf{Online Demo:} \href{https://www.cpret.online/}{https://www.cpret.online/}
\end{center}

\section{Introduction}
Competitive programming contests—from high-school Olympiads like the IOI to university-level events such as the ICPC—challenge participants to solve algorithmic problems under tight time and memory limits, requiring both strong coding skills and deep algorithmic insight. Because problem statements are precisely specified in natural language and solutions can be graded automatically, these problems have become a canonical benchmark for assessing the reasoning and coding abilities of large language models (LLMs) \cite{livecodebench, humaneval}. Recent advances in code-oriented LLMs—spanning commercial flagships such as OpenAI o4-mini \cite{o3_o4mini}, Gemini-2.5-Pro \cite{gemini25pro}, and Grok-3-Mini (High) \cite{grok3}, alongside research releases like DeepSeek-R1 \cite{deepseek-r1} highlight the value of competitive programming as a testbed for algorithmic reasoning, program synthesis, and computer-science education.

\begin{wrapfigure}{r}{0.5\linewidth}
    \centering
    \includegraphics[width=\linewidth]{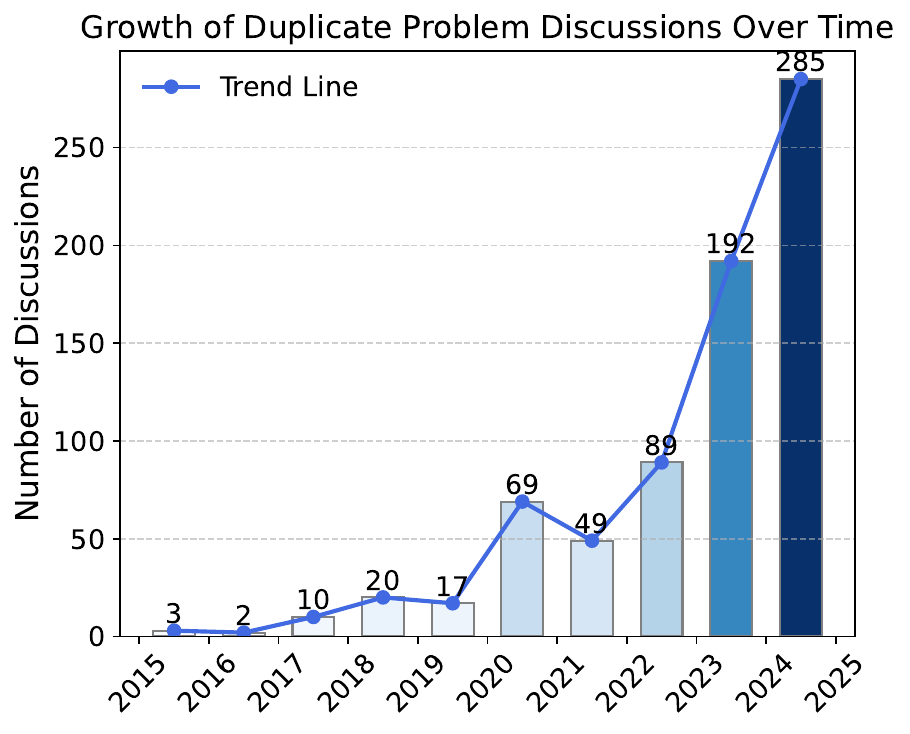}
    \caption{Year-wise trend of annotated duplicate problem discussions. 
    }
    \label{fig:trend-duplicate-discussions}
\end{wrapfigure}

Competitive programming has witnessed rapid growth over the past three decades, with thousands of new problems introduced annually. This expansion has led to a growing concern: the increasing presence of similar or repetitive problems within large repositories. As shown in Figure~\ref{fig:trend-duplicate-discussions}, our collected data reveals a significant rise in community discussions around duplicate problems in recent years.  However, it is often difficult to determine whether a new problem is a duplicate of existing ones, as this typically relies on the memory and judgment of human problem setters. Unchecked duplication brings concrete downsides. In human programming contests, repeated or highly similar problems give an unfair advantage to participants who have seen them before. In competitive programming benchmarks for LLMs, the presence of repeated or highly similar problems can lead to inflated performance scores, as models may rely on memorized instances of past competition problems from their training data rather than demonstrating genuine algorithmic reasoning. This compromises the evaluation of their true ability to solve novel and unfamiliar challenges.

Competitive programming problem retrieval poses unique challenges compared to standard code retrieval: problem statements are often abstract, narrative-driven, and may admit multiple fundamentally different solutions, requiring models to capture high-level algorithmic ideas rather than surface code patterns. This motivates the need for specialized problem-level retrieval models.

A promising way to address the issue of duplicate problems is to leverage retrieval models that identify semantically similar problems based on learned representations. Such models offer a scalable and model-agnostic solution for redundancy detection in large problem repositories. However, despite their potential, there is a lack of established benchmarks specifically targeting problem-level retrieval. Prior efforts in competitive programming retrieval have primarily focused on code-centric tasks, such as retrieving solution code given a problem description (\emph{Text-to-Code}) or retrieving alternative correct solutions given a reference implementation (\emph{Code-to-Code}), which overlook the dimension of retrieving similar problems themselves—an essential capability for both redundancy detection and educational search.

To fill this gap, we introduce two new problem-centric retrieval tasks. (I) \emph{Problem-to-Duplicate} asks a model to retrieve the most semantically or structurally similar problem to a given target. We manually annotated 700 duplicate pairs; 200 pairs form a held-out test set in which the model must locate the duplicate among a corpus of 10,000+ problems, and the remaining $\sim 500$ pairs are used for training. (II) \emph{Simplified-to-Full} supplies a simplified or paraphrased statement and requires the model to find the original full version. We collected 17,000 human-written simplifications aligned with their source problems; 10,000 pairs constitute the test set and the remaining $\sim 7,000$ pairs serve as training data. These tasks target practical problem-retrieval needs—spotting duplicate or highly similar problems and retrieving the original statement from a simplified description. In addition, we perform a temporal analysis of the two code-centric tasks and uncover severe train-test leakage: models fare markedly worse on the newest problems. To this end, we rebuild both benchmarks, drawing the test sets exclusively from the most recent problems to enforce strict temporal separation.

Furthermore, to address the real-world retrieval demands, we develop two task-specialized models CPRetriever-Code and CPRetriever-Prob using our new corpus: (I) the first one is trained on 38 k problems and 2.9 M multi-language solutions with a new \emph{Group-InfoNCE} loss that treats every correct solution of the same problem as a positive and penalizes similarity within that positive set. This encourages alignment across diverse implementations, and it achieves the overall best performance on all code-centric tasks. (II) The second one is fine-tuned from CPRetriever-Code on our duplicate-detection and simplified-to-full datasets, and it performs the overall best results on the two problem-centric tasks.

Finally, we apply CPRetriever-Prob to analyze problems in LiveCodeBench released after September 2024 by computing each problem’s maximum similarity to earlier data. We observe two key trends: (I) \emph{Pass rates increase with similarity}—problems more similar to prior examples are consistently easier for models to solve, regardless of difficulty; (II) \emph{High similarity reduces model differentiation}—as similarity grows, performance gaps between models shrink, suggesting that models may solve such problems by relying on memorized patterns rather than demonstrating true reasoning ability in novel scenarios. These findings highlight the importance of accounting for problem similarity when constructing benchmarks to avoid overestimating model capabilities.

Overall, our key contributions can be summarized as follows:

$\bullet$ \textbf{Comprehensive Benchmark:} We construct a large-scale, high-quality dataset for competitive programming, covering four key retrieval tasks (two focused on code retrieval and two newly proposed for problem retrieval). Our benchmark reflects realistic retrieval needs in programming contests and addresses issues such as problem repetition and dataset leakage.
    
$\bullet$ \textbf{Effective Models:} We develop two task-specialized retrieval models with strong performance, trained with our newly designed Group-InfoNCE loss. Both models are open-sourced and can be run locally—an important feature given the privacy constraints in contest settings—providing practical tools for code and problem retrieval and helping mitigate the growing issue of duplicate problems in competitive programming.

$\bullet$ \textbf{Similarity-Aware Evaluation:} We provide insightful studies on the impact of problem similarity in competitive programming, and find that problems with high similarity to prior data yield inflated pass rates and reduced model differentiation, especially for weaker models. These results highlight the importance of controlling for similarity when designing benchmarks and suggest that future datasets should stratify or filter test problems based on their similarity to past content.


\section{Related Work}

\subsection{Datasets for Competitive Programming}

Several datasets have been proposed to train and evaluate LLMs on competitive programming tasks. Early efforts include Description2Code \cite{Description2Code}, which gathered problems and solutions from CodeChef and Codeforces, followed by the APPS dataset \cite{apps}, which became a widely used benchmark across a mix of competition-style and introductory problems. CodeContests \cite{alphacode}, used in DeepMind’s AlphaCode, expanded coverage but remained limited in scale. More recently, TACO \cite{TACO} constructed the largest dataset (26K problems) by aggregating multiple sources. In the retrieval setting, CoIR \cite{CoIR} built on APPS and CodeTransOcean \cite{codetransocean} to define two retrieval tasks: text-to-code and code-to-code, helping assess model capabilities in aligning problem descriptions and solution code. Despite their contributions, these datasets have several limitations: (i) many were collected before 2023 and suffer from potential data leakage, (ii) problem types are often narrow in scope, and (iii) both problem statements and solutions are typically limited to English and Python, reducing linguistic and implementation diversity.

For evaluating programming ability, HumanEval \cite{humaneval} remains a widely used standard, featuring hand-written Python problems with corresponding unit tests for functional correctness. LiveCodeBench \cite{livecodebench} improves upon earlier efforts by continuously collecting real-world problems from LeetCode, AtCoder, and Codeforces, and applies temporal separation to reduce data contamination. However, these benchmarks do not account for problem-level similarity between new and historical problems—a distinct issue from data leakage—which can still lead to overly optimistic assessments of model performance due to hidden redundancy.


\subsection{Embedding Models for Retrieval}

Dense retrieval models learn to map queries and candidates into a shared embedding space for efficient similarity search. Early methods like DSSM \cite{dssm} and Sentence-BERT \cite{sentencebert} paved the way for supervised contrastive training, leading to influential retrievers such as Dense Passage Retrieval (DPR) \cite{dpr}.

Recent efforts have focused on building general-purpose embedding models (e.g., E5-Mistral \cite{e5mistral}, GTE-Qwen2 \cite{gteqwen2}, SFR-Embedding \cite{sfrtext}, NV-Embed \cite{nvembed,nv-retriever}, Linq-Embed\cite{Linq}) that perform well across a wide range of retrieval tasks. These models are commonly evaluated on MTEB \cite{MTEB}, a standardized benchmark suite covering diverse tasks such as semantic search, classification, and reranking. For the code domain, specialized models like CodeSage \cite{codesage}, SFR-Embedding-Code \cite{sfrcode}, Qodo-Embed\cite{qodo} use code-specific data and retrieval-aware objectives to significantly improve performance on benchmarks such as CoIR. While both general-purpose and code-specific embedding models can be applied to competitive programming, we find their performance on code-centric and problem-centric retrieval tasks remains limited—highlighting the need for a dedicated solution tailored to the unique characteristics of this domain.

High-quality retrieval has been shown to benefit downstream tasks like code generation via Retrieval-Augmented Generation (RAG), where an LLM leverages retrieved problem-solution pairs as context. For example, Shi et al.~\cite{shi2024can} explored solving Olympiad-level problems with RAG, Li et al.~\cite{li2024can} proposed critic-guided retrieval-augmented planning, and Mapcoder~\cite{islam2024mapcoder} introduced MapCoder, a multi-agent code generation framework using retrieval. These studies suggest that problem-level retrieval can substantially enhance LLM performance, motivating specialized embeddings for competitive programming.


While temporal splits and duplicate-problem detection are not new concepts—having been explored in other domains such as finance~\cite{de2018advances} and software community question answering~\cite{hazra2023duplicate}—their systematic application to competitive programming remains underexplored. Existing benchmarks (e.g., MTEB\cite{MTEB}, CoIR\cite{CoIR}) often lack strict temporal separation, potentially inflating evaluation results. Our work extends these established principles to programming contests, where duplicate or temporally overlapping problems introduce unique challenges: algorithmically similar tasks may differ superficially in text but still bias performance. By constructing a benchmark with rigorous temporal splits and fine-grained similarity filtering, we aim to build fairer evaluation standards and reveal new insights into the relationship between problem similarity and model discrimination ability.
\section{Method}

\subsection{Overview of Retrieval Tasks}

\label{sec:retrieval_tasks}
We define and evaluate four distinct retrieval tasks, each capturing a different dimension of semantic understanding and reuse in competitive programming:

\begin{itemize}
\item \textbf{Text-to-Code Retrieval}: Given the full natural language description of a problem, retrieve one or more correct solution codes. This task assesses a model’s ability to align problem semantics with executable implementations.

\item \textbf{Code-to-Code Retrieval}: Given one accepted solution, retrieve alternative correct solutions to the same problem. This task evaluates a model’s capacity to understand code functionality and identify semantically equivalent but syntactically diverse implementations.

\item \textbf{Problem-to-Duplicate Retrieval}: Given a problem description, retrieve other problems that are duplicated, including exact matches or closely related in terms of solution strategy. This task is useful for identifying redundancy and measuring problem novelty across platforms.

\item \textbf{Simplified-to-Full Retrieval}: Given a simplified version of a problem, retrieve the corresponding one with full description. This task examines cross-abstraction retrieval capabilities, bridging accessibility-oriented rewrites and original problem statements.
\end{itemize}

These tasks jointly form a comprehensive benchmark for evaluating retrieval models in the context of competitive programming, covering problem-code alignment, solution diversity, duplication detection, and abstraction-level matching. Table~\ref{tab:task-statistics} summarizes the statistics of each retrieval task, including training and test set sizes, as well as average token lengths for both queries and corpus items.

\begin{table}[h]
\centering
\caption{\textbf{Statistics of the four retrieval tasks in CPRet.} 
We report the number of training and test items, where \#Train-Code indicates the number of distinct solution codes, and \#Train-Pair refers to the number of (anchor, positive) pairs used for training. 
$L_{\text{Query}}$ and $L_{\text{Corpus}}$ denote average token lengths.}
\label{tab:task-statistics}
\resizebox{0.9\linewidth}{!}{
\begin{tabular}{c@{}cccccccc}
\toprule
\shortstack[c]{\raisebox{0.8ex}{Retrieval Task}} 
& \shortstack[c]{\#Train-\\Problem} 
& \shortstack[c]{\#Train-\\Code} 
& \shortstack[c]{\#Train-\\Pair} 
& \shortstack[c]{\#Test-\\Query} 
& \shortstack[c]{\#Test-\\Corpus} 
& \shortstack[c]{\#Test-\\Qrels} 
& \raisebox{0.8ex}{$L_{\text{Query}}$} 
& \raisebox{0.8ex}{$L_{\text{Corpus}}$} \\
\midrule
\shortstack[c]{Text-to-Code} 
  & \multirow{2}{*}{38.8K} 
  & \multirow{2}{*}{2.93M} 
  & \multirow{2}{*}{2.93M} 
  & 4.9k & 41.6k & 41.6k & 1038 & 1210 \\
\cmidrule(lr){5-9}
\shortstack[c]{Code-to-Code} 
  & & & & 4.8k & 39.8k & 39.8k & 1132 & 1185 \\
\midrule
\shortstack[c]{Problem-to-Duplicate} & 874 & / & 491 & 168 & 10.9k & 202 & 565 & 765 \\
\midrule
\shortstack[c]{Simplified-to-Full}     & 7.6k & / & 7.6k & 10k & 10k & 10k & 226 & 697 \\
\bottomrule
\end{tabular}}
\end{table}

\subsection{Dataset and Benchmark Construction}
\subsubsection{Problems and Codes Data}
\label{sec:leakage}
We construct a large-scale dataset of programming contest problems and their corresponding accepted solutions to support both model training and the evaluation of retrieval tasks. Compared to existing datasets, our dataset provides \emph{broader temporal coverage}, \emph{richer language diversity}, and \emph{more varied contest formats}. It includes recent problems from multiple online judges, spans several programming languages, and covers both ICPC-style (International Collegiate Programming Contest) and OI-style (Olympiad in Informatics) problems—the latter involving partial scoring and more complex algorithmic requirements. Each problem is paired with one or more solutions and annotated with precise timestamp information, enabling temporally-aware training and evaluation across semantic retrieval tasks. As summarized in Table~\ref{tab:cp_datasets}, our dataset—\textbf{CPRet-PCPCD}—offers wider coverage across problem sources, types, and languages, and serves as a stronger foundation for retrieval-based modeling in competitive programming.

\begin{table}[h]
\centering
\caption{
\textbf{Comparison of competitive programming datasets. }
\textbf{\#Src}: Number of data sources ({\it e.g.}, online judges). 
\textbf{\#P-Type}: Number of problem types ({\it e.g.}, ICPC-style with full-score only vs OI-style with partial scoring). 
\textbf{\#Lang-P}: Number of languages used for problem descriptions. 
\textbf{\#Lang-C}: Number of programming languages used in solutions. See Appendix~\ref{appendix:dataset_stats} for more detailed dataset statistics.
}
\label{tab:cp_datasets}
\resizebox{0.9\linewidth}{!}{
\begin{tabular}{lrrcccccc}
\toprule
Dataset & \#Prob & \#Code & \#Src & \#P-Type & \#Lang-P & \#Lang-C & Cut-off \\
\midrule
Description2Code \cite{Description2Code} & 7.8K  & 309K  & 3  & 1 & 1 & 2   & 2016/08 \\
APPS\cite{apps}      & 10K   & 232K  & 7  & 1 & 1 & 1   & 2020/10 \\
CodeContests\cite{alphacode}  & 13.6K & 4.5M  & 5  & 1 & 1 & 4+  & 2021/07 \\
TACO\cite{TACO}      & 26.4K & 1.55M & 10 & 1 & 1 & 1   & 2023/02 \\
\textbf{CPRet-PCPCD (ours)}     & \textbf{42.2K} & \textbf{2.9M}  & \textbf{12} & \textbf{2} & \textbf{3} & \textbf{20+} & \textbf{2024/12} \\
\bottomrule
\end{tabular}}
\end{table}

\begin{figure}[t]
    \centering
    \begin{minipage}[t]{0.49\linewidth}
        \centering
        \includegraphics[width=\linewidth]{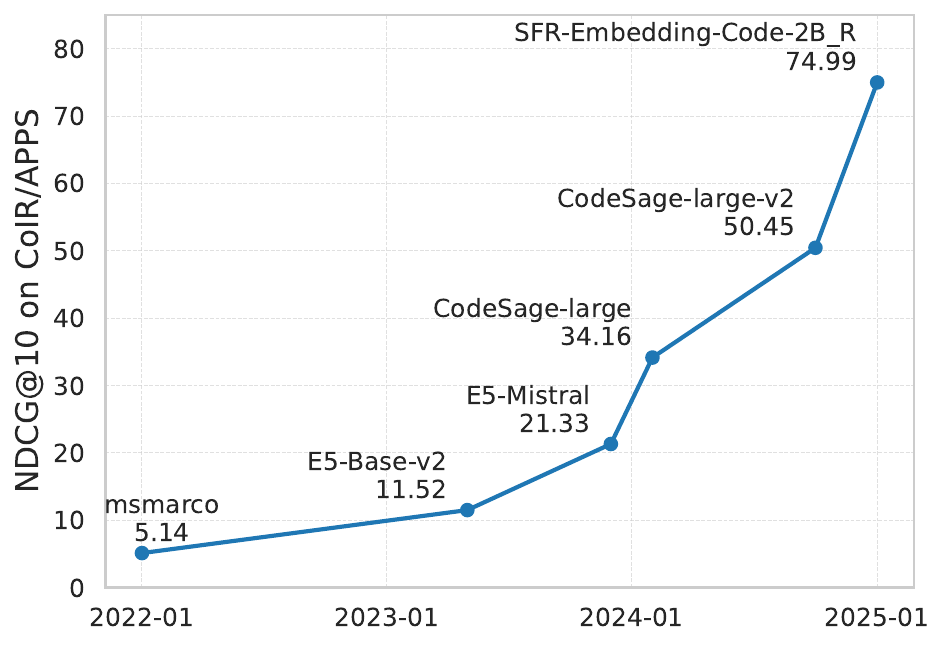}
    \end{minipage}
    \hfill
    \begin{minipage}[t]{0.49\linewidth}
        \centering
        \includegraphics[width=\linewidth]{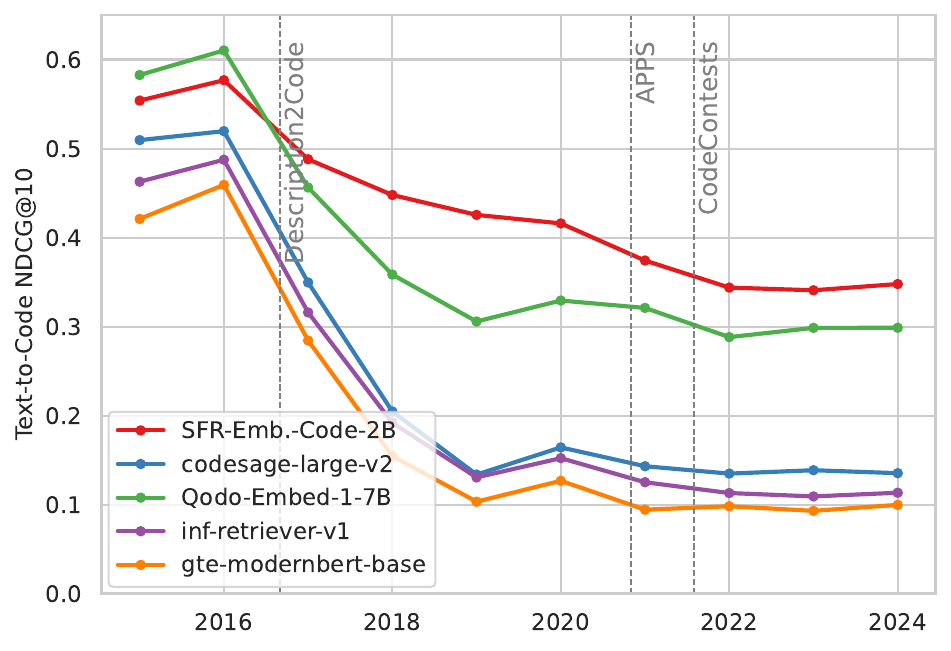}
    \end{minipage}
    \caption{
        \textbf{Left:} NDCG@10 scores on CoIR/APPS across model release dates, showing steady performance improvement as models evolve. 
        \textbf{Right:} Performance on the Text-to-Code task grouped by problem release year, revealing sharp degradation on older problems, especially around major dataset release points (\textit{Description2Code}, \textit{APPS}, \textit{CodeContests}).
    }
    \label{fig:temporal-analysis}
\end{figure}

As shown in Figure \ref{fig:temporal-analysis}, on the Text-to-Code task using the historical APPS dataset, model performance improves over time as newer and stronger models are introduced. However, we observe a clear drop in performance when problems are grouped by their original release year—only stabilizing after 2022. This indicates potential data leakage or memorization in earlier benchmarks and underscores the need for temporally separated evaluation. Accordingly, we use problems and solutions from 2023 onward in our collected dataset as the test set for both the Text-to-Code and Code-to-Code tasks.

\subsubsection{Duplicate Programming Problem Pairs}

In many instances, when a newly released contest problem closely resembles or duplicates a previously published one, participants will highlight this in the associated discussion threads or contest forums. To leverage this observation, we collected all publicly available discussion threads and blog posts from two major competitive programming platforms: Codeforces and Luogu. We applied a combination of keyword-based heuristics and large language model (LLM)-based classification to identify approximately 5,000 potentially relevant entries. These candidates were subsequently verified through manual annotation by several experienced competitive programmers.

To ensure consistency and clarity in annotation, we defined three levels of duplication between problems:

\begin{itemize}
    \item Exact Match: An accepted (AC) solution for one problem can directly pass the other without modification.
    \item Near Match: An AC solution for one problem can be adapted to solve the other with minor edits that could reasonably be made by someone unfamiliar with the second problem.
    \item Method Match: The core idea and solution approach are the same, but the code differs in non-essential or implementation-specific details.
\end{itemize}

We ultimately identified around 700 pairs of duplicate problems. Since some problems may belong to duplication clusters involving more than two problems, we first performed clustering to group mutually duplicate problems. We then randomly selected 30\% of the clusters to construct the test set for the \textbf{Problem-to-Duplicate} retrieval task. Within each selected cluster, one problem was randomly designated as the query, while the remaining ones were placed in the corpus. To further increase task difficulty and realism, we added all other Codeforces problems to the corpus as distractors.

\subsubsection{Simplified and Full Problem Description Pairs}

To support beginners and non-native speakers, users on the Luogu platform have contributed Chinese translations and simplified versions of competitive programming problems originally written in English ({\it e.g.}, on Codeforces) or Japanese ({\it e.g.}, on AtCoder). We crawled these user-generated simplifications and applied filtering procedures to remove low-quality entries, particularly those stemming from direct or unedited machine translation.

After cleaning, we obtained approximately 17,000 high-quality pairs of full problem descriptions and their corresponding simplified versions. From this set, we randomly selected 10,000 pairs to construct the test set for the \textbf{Simplified-to-Full} retrieval task, where the simplified description serves as the query and the full version is used as the corpus. The remaining around 7,000 pairs are used for training.

\subsection{Competitive Programming Retrievers}

\subsubsection{CPRetriever-Code: Multi-Positive Contrastive Learning for Problem-Code Alignment}

We train CPRetriever-Code using a supervised contrastive learning framework tailored for aligning problem descriptions with diverse correct solutions. Each problem is paired with multiple valid codes, and the model learns to capture this alignment while preserving representation diversity.

We begin with the standard InfoNCE loss~\cite{oord2018representation}, where each (problem, solution) pair $(x_i, x_i^+)$ is contrasted against others in the batch:
\begin{equation}
\mathcal{L}_{\text{InfoNCE}} = -\log \frac{\exp(\mathrm{sim}(x_i, x_i^+)/\tau)}{\sum_{j=1}^{N} \exp(\mathrm{sim}(x_i, x_j^+)/\tau)}.
\label{eq:infoNCE}
\end{equation}

\noindent
\textbf{Limitation of Single-Positive Loss.}
In practice, many problems have multiple correct solutions $G_i = {x_i^{1+}, \ldots, x_i^{m+}}$. A common extension averages the similarities over all positives~\cite{zhao2023leveraging}:
\begin{equation}
\mathcal{L}_{\text{MultiPos}} = -\log \frac{\sum_{k=1}^{m} \exp\left(\mathrm{sim}(x_i, x_i^{k+}) / \tau\right)}{\sum_{j \ne i}^{N} \exp\left(\mathrm{sim}(x_i, x_j) / \tau\right)}.
\label{eq:multipos}
\end{equation}
However, this ignores the internal structure of the positive set and does not encourage consistency among the positives.

\noindent
\textbf{Group-InfoNCE.}
To better utilize multiple correct solutions, we propose Group-InfoNCE, which treats the positive set as a whole. The loss promotes similarity between the query $x_i$ and its group $G_i$, while contrasting against other problems and groups:

\begin{equation}
\begin{aligned}
\mathcal{L}_{\text{Group}} =\; & -\log \frac{\exp\left(\mathrm{sim}_G(x_i, G_i)/\tau\right)}{
\exp\left(\mathrm{sim}_G(x_i, G_i)/\tau\right) 
+ \sum_{j \neq i} \left[
\exp\left(\mathrm{sim}(x_i, x_j)/\tau\right) + 
\exp\left(\mathrm{sim}_G(x_i, G_j)/\tau\right)
\right]} \\
& + \frac{\text{Penalty}_G(x_i, G_i)}{\tau^2}.
\end{aligned}
\label{eq:groupinfonce}
\end{equation}

Here, group similarity is defined as:
\begin{equation}
\mathrm{sim}_G(x_i, G_j) = \frac{1}{m} \sum_{k=1}^{m} \mathrm{sim}(x_i, x_j^{k+}),
\label{eq:group_sim}
\end{equation}
and the variance-based regularization encourages consistency within the group:
\begin{equation}
\text{Penalty}_G(x_i, G_i) = \mathrm{Var}_{k=1}^m \left( \mathrm{sim}(x_i, x_i^{k+}) \right).
\label{eq:group_penalty}
\end{equation}

This group-based formulation improves representation quality by explicitly modeling the structure of multiple correct solutions, resulting in more stable and discriminative embeddings for code-centric tasks. Although Group-InfoNCE enforces consistency between a problem embedding and its set of correct solutions, it does not constrain the diversity of the solutions themselves. The variance-based regularization ensures that each solution maintains a consistent similarity to the problem embedding, effectively "surrounding" it in the code space without collapsing distinct approaches. In practice, these similarities converge around an average value, allowing the problem representation to capture the essential idea while accommodating diverse algorithmic strategies. Empirical results on both base models (see Tables~\ref{tab:stage1-ablation} and \ref{tab:qwen4b-multitask}) show consistent improvements in retrieval performance and robust, discriminative embeddings.

\noindent
{\bf Format Masking for Robustness.}
To reduce reliance on superficial cues, we apply random masking to non-essential parts of the problem description during training, including I/O format explanations, sample inputs/outputs, and explicit data constraints. This encourages the model to focus on the core algorithmic intent rather than overfitting to formatting patterns, and improves generalization to problems with diverse or unfamiliar layouts.

\subsubsection{CPRetriever-Prob: Fine-Tuning on Problem-Level Tasks}

CPRetriever-Prob is obtained by fine-tuning CPRetriever-Code on problem-level retrieval tasks. While CPRetriever-Code captures general alignment between problems and solutions, this stage specializes the model for retrieving semantically or structurally related problems—such as duplicates or simplified versions.

In particular, we jointly train on data from the \textit{Problem-to-Duplicate} and \textit{Simplified-to-Full} tasks (Section~\ref{sec:retrieval_tasks}), enabling the model to learn fine-grained alignment between problems that are either functionally equivalent or differ primarily in abstraction level.

\noindent
\textbf{Training Format and Loss.}
Each training example is structured as a triplet $(x, x^+, x^-)$, where $x$ is the query, $x^+$ is a semantically aligned problem, and $x^-$ is a hard negative. We use the \textit{triplet margin loss}~\cite{schroff2015facenet}:
\begin{equation}
\mathcal{L}_{\text{triplet}} = \max\left(0, \mathrm{sim}(x, x^-) - \mathrm{sim}(x, x^+) + \alpha\right),
\label{eq:triplet_loss}
\end{equation}
where $\mathrm{sim}(\cdot, \cdot)$ is cosine similarity and $\alpha$ is the margin.

\noindent
\textbf{Hard Negative Mining.}
As our dataset primarily contains positive pairs, we construct hard negatives by first retrieving the top-10 most similar candidates for each query using a pretrained retriever (\texttt{SFR-Embedding-Code-2B\_R}\cite{sfrcode}) and randomly sampling one as a negative. 
To ensure the sampled negative is not a false match, we use \texttt{Qwen-2.5-Max}\cite{qwen2} to verify that it is not equivalent to the query in terms of problem intent. This automatic filtering yields challenging yet reliable negatives without requiring manual annotation.
\section{Experiment}

\subsection{Experimental Setup}

We develop two retrieval models based on the \texttt{SFR-Embedding-Code-2B\_R}\cite{sfrcode} embedding model, which is built on top of \texttt{Gemma-2-2B}\cite{gemma} and further pre-trained on code-related tasks. We further evaluate our method on the \texttt{Qwen3-Embedding-4B}\cite{qwen3emb} model, released in June 2025, which demonstrates stronger retrieval capabilities. Detailed training configurations are provided in Appendix~\ref{appendix:exp_setup}.

\subsection{Main Results}

\begin{table*}[htbp]
\centering

\caption{
Retrieval performance across models with different scales. 
We use NDCG@10~\cite{jarvelin2002cumulated} as the primary metric to measure ranking quality across tasks. 
\textbf{T2C}: Text-to-Code. 
\textbf{C2C}: Code-to-Code. 
\textbf{P2Dup}: Problem-to-Duplicate. 
\textbf{S2Full}: Simplified-to-Full. 
\textbf{Avg}: Average of the four tasks.
}
\resizebox{0.9\linewidth}{!}{
\renewcommand{\arraystretch}{1.1}
\begin{tabularx}{\textwidth}{c X l l c c c c c}
\toprule
\textbf{Scale} & \textbf{Model} & \textbf{Type} & \textbf{Size} & \textbf{T2C} & \textbf{C2C} & \textbf{P2Dup} & \textbf{S2Full} & \textbf{Avg} \\
\midrule

\multirow{4}{*}{Tiny} 
& {\scriptsize contriever-msmarco~\cite{contriever-msmarco}} & general & 109M & 1.06 & 7.49 & 5.26 & 43.04 & 14.21 \\
& {\scriptsize multilingual-e5-small~\cite{me5}} & general & 110M & 2.73 & 13.52 & 14.90 & 52.07 & 20.80 \\
& {\scriptsize codesage-small-v2~\cite{codesage}} & code & 130M & 10.16 & 22.24 & 17.42 & 67.07 & 29.22 \\
& {\scriptsize gte-modernbert-base~\cite{mgte}} & general & 149M & 14.99 & 36.22 & 21.12 & 77.45 & 37.44 \\
\midrule

\multirow{9}{*}{Small}
& {\scriptsize bge-large-zh-v1.5~\cite{bge}} & general & 324M & 2.58 & 7.49 & 14.75 & 37.34 & 15.54 \\
& {\scriptsize stella\_en\_400M\_v5~\cite{stella}} & general & 400M & 2.12 & 9.76 & 13.14 & 57.56 & 20.64 \\
& {\scriptsize multilingual-e5-base~\cite{me5}} & general & 278M & 1.96 & 13.66 & 17.24 & 58.26 & 22.78 \\
& {\scriptsize bge-m3~\cite{bge-m3}} & general & 569M & 5.69 & 12.67 & 20.12 & 63.81 & 25.57 \\
& {\scriptsize codesage-base-v2~\cite{codesage}} & code & 356M & 17.30 & 29.38 & 19.97 & 74.36 & 35.25 \\
& {\scriptsize SFR-Emb.-Code-400M~\cite{sfrcode}} & code & 400M & 9.43 & 43.59 & 19.40 & 75.31 & 36.93 \\
& {\scriptsize multilingual-e5-large~\cite{me5}} & general & 560M & 4.27 & 18.51 & 19.19 & 65.02 & 26.75 \\
& {\scriptsize multi.-e5-large-instruct~\cite{me5}} & general & 560M & 6.64 & 28.84 & 23.31 & 61.28 & 30.02 \\
& {\tiny Qwen3-Embedding-0.6B~\cite{qwen3emb}} & general & 600M & 48.96 & 60.49 & 36.26 & 81.63 & 56.83 \\
\midrule

\multirow{12}{*}{Medium}
& {\scriptsize gte-Qwen2-1.5B-inst.~\cite{gte}} & general & 1.5B & 10.76 & 23.41 & 27.06 & 69.15 & 32.60 \\
& {\scriptsize stella\_en\_1.5B\_v5~\cite{stella}} & general & 1.5B & 9.22 & 21.40 & 29.45 & 72.91 & 33.24 \\
& {\scriptsize inf-retriever-v1-1.5b~\cite{inf}} & general & 1.5B & 18.31 & 28.17 & 30.11 & 74.19 & 37.70 \\
& {\scriptsize codesage-large-v2~\cite{codesage}} & code & 1.3B & 20.78 & 35.23 & 22.43 & 78.70 & 39.28 \\
& {\scriptsize Qodo-Embed-1-1.5B~\cite{qodo}} & code & 1.5B & 22.93 & 36.52 & 33.37 & 84.05 & 44.22 \\
& {\scriptsize SFR-Embedding-2~\cite{sfrtext}} & general & 2B & 20.57 & 50.38 & 35.96 & 73.02 & 44.98 \\
& {\scriptsize SFR-Emb.-Code-2B~\cite{sfrcode}} & code & 2B & 39.60 & 68.05 & 45.26 & 86.43 & 59.84 \\
\rowcolor{yellow!20}
& {\scriptsize CPRetriever-Code} & code & 2B & 70.40 & 70.59 & 38.68 & 81.45 & 65.28 \\
\rowcolor{yellow!20}
& {\scriptsize CPRetriever-Prob} & code & 2B & 56.50 & 70.68 & 60.06 & 90.74
 & 69.50 \\
 & {\scriptsize Qwen3-Embedding-4B~\cite{qwen3emb}} & general & 4B & 66.62 & 71.97 & 56.59 & 89.39 & 71.15 \\
 \rowcolor{yellow!20}
& {\tiny CPRetriever-Code-Qwen3-4B} & code & 4B & \textbf{86.22} & 86.70 & 41.14 & 88.10 & 75.54 \\
\rowcolor{yellow!20}
& {\tiny CPRetriever-Prob-Qwen3-4B} & code & 4B & 80.84 & \textbf{87.10} & \textbf{74.33} & \textbf{96.15} & \textbf{84.60} \\

\midrule

\multirow{8}{*}{Large}
& {\scriptsize GritLM-7B~\cite{Grit}} & general & 7B & 0.22 & 8.74 & 11.18 & 29.81 & 12.49 \\
& {\scriptsize NV-Embed-v2~\cite{nv-retriever,nvembed}} & general & 7B & 7.09 & 34.88 & 25.49 & 63.14 & 32.65 \\
& {\scriptsize inf-retriever-v1~\cite{inf}} & general & 7B & 17.43 & 25.38 & 30.85 & 78.46 & 38.03 \\
& {\scriptsize gte-Qwen2-7B-instruct~\cite{gte}} & general & 7B & 17.96 & 30.72 & 35.95 & 78.41 & 40.76 \\
& {\scriptsize SFR-Emb.-Mistral~\cite{sfrtext}} & general & 7B & 22.15 & 50.92 & 31.88 & 69.38 & 43.58 \\
& {\scriptsize Linq-Embed-Mistral~\cite{Linq}} & general & 7B & 21.99 & 52.06 & 36.51 & 72.79 & 45.84 \\
& {\scriptsize Qodo-Embed-1-7B~\cite{qodo}} & code & 7B & 36.47 & 51.91 & 47.15 & 91.17 & 56.68 \\
& {\scriptsize Qwen3-Embedding-8B~\cite{qwen3emb}} & general & 8B & 60.54 & 72.97 & 53.23 & 87.95 & 68.67 \\
\bottomrule
\label{tab:model_comparison}
\end{tabularx}}
\vspace{-20pt}
\end{table*}

\noindent

We evaluate over 20 strong embedding models drawn from top-performing entries on the MTEB\cite{MTEB} and CoIR\cite{CoIR} benchmarks, as summarized in Table~\ref{tab:model_comparison}. Models trained specifically for the code domain continue to outperform general-purpose models by a significant margin. Among them, our two proposed models—\textbf{CPRetriever-Code} and \textbf{CPRetriever-Prob}—achieve the best overall results across tasks.

\textbf{CPRetriever-Code}, trained with Group-InfoNCE on problem-code pairs, excels on the code-centric tasks—Text-to-Code (T2C) and Code-to-Code (C2C). After fine-tuning on problem-level data, \textbf{CPRetriever-Prob} achieves large gains on the problem-centric tasks—Problem-to-Duplicate (P2Dup) and Simplified-to-Full (S2Full)—with a modest drop on C2C and a more noticeable decline on T2C.

We attribute the T2C degradation to differing retrieval demands: T2C relies on implementation-specific details in problem descriptions, whereas P2Dup and S2Full emphasize higher-level semantic similarity. This trade-off is also evident in other models—for instance, \textit{Qodo-Embed-1-7B}\cite{qodo} achieves the best S2Full score but underperforms on T2C compared to \textit{SFR-Embedding-Code-2B\_R}\cite{sfrcode}. These observations motivate our decision to release two task-specialized models, each optimized for different objectives. Further analysis of this trade-off is provided in Appendix~\ref{stage2:ablation}, and detailed ablation studies on the Group-InfoNCE loss appear in Appendix~\ref{stage1:ablation}.

\subsection{Impact of Problem Similarity on Model Success in Competitive Programming Tasks}

To assess how prior problem similarity influences model performance, we examine 388 LiveCodeBench problems released after September 1, 2024. For each, we compute the average pass rate across all evaluated models and measure its maximum cosine similarity to earlier CPRet-PCPCD problems using \textbf{CPRetriever-Prob}.

\begin{figure}[t]
    \centering
    \begin{minipage}[t]{0.49\linewidth}
        \centering
        \includegraphics[width=\linewidth]{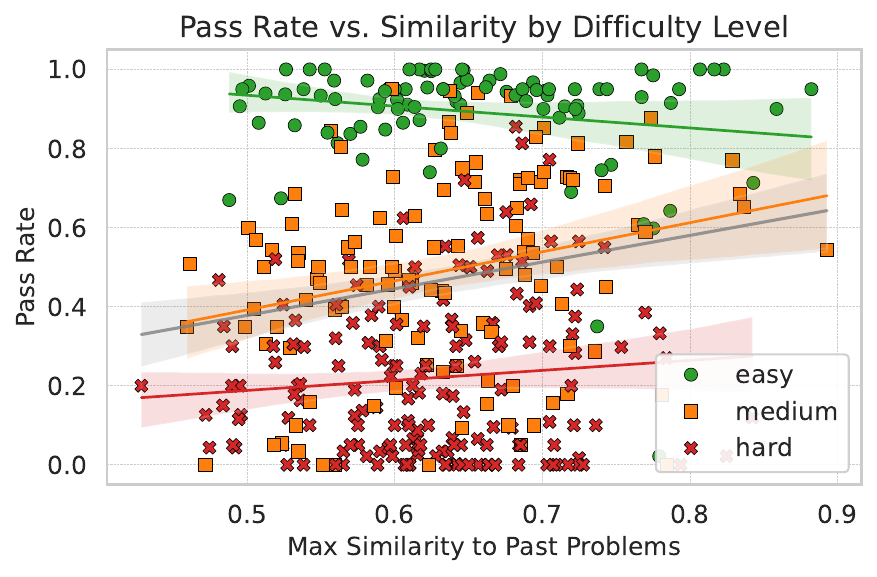}
    \end{minipage}
    \hfill
    \begin{minipage}[t]{0.49\linewidth}
        \centering
        \includegraphics[width=\linewidth]{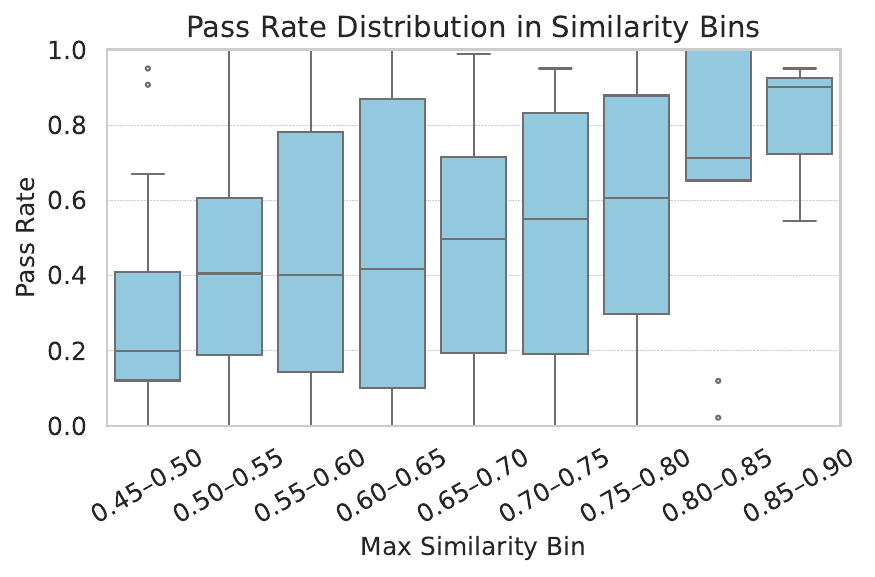}
    \end{minipage}
    \vspace{-3pt}
    \caption{
        \textbf{Impact of similarity to prior problems on pass rate.} 
        \textbf{Left:} For 388 post-2024.9.1 LiveCodeBench problems, we plot average model pass rate vs. maximum similarity to historical CPRet problems, computed via \textbf{CPRetriever-Prob}. Regression lines are shown by difficulty, with the gray line indicating the overall trend across all problems.
        \textbf{Right:} Pass rates increase monotonically across similarity bins, confirming a strong link between retrieval similarity and generation success.
    }
    \label{fig:sim-passrate}
    \vspace{-10pt}
\end{figure}

As shown in Figure~\ref{fig:sim-passrate}, we observe a clear trend: on average, problems with higher maximum similarity to past data have significantly higher pass rates. This pattern holds across difficulty levels, but with notable differences:
(I) Medium-difficulty problems show the strongest positive correlation, suggesting that models benefit substantially from prior exposure to semantically similar content. (II) Hard problems exhibit generally low pass rates, though a slight upward trend remains visible with increasing similarity. (III) Easy problems maintain consistently high pass rates overall. Due to a few low-performing outliers at high similarity, the regression line shows a slight downward slope—likely an artifact of skewed distribution rather than a true negative effect. The binned box plot (right) further confirms this relationship, reinforcing that semantic similarity to known problems is a key factor in model success.

\begin{figure}[t]
    \centering
    \begin{minipage}[t]{0.57\linewidth}
        \centering
    \includegraphics[width=0.9\linewidth]{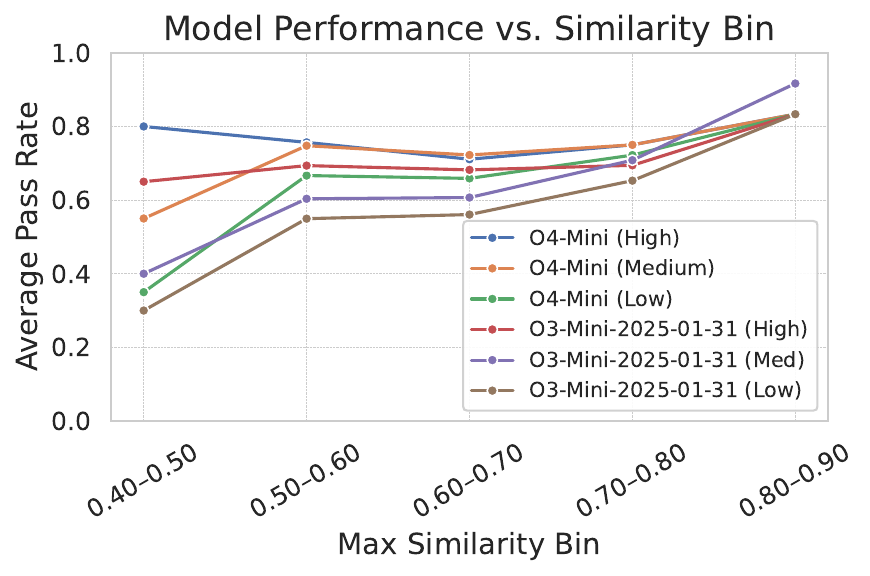}
    \end{minipage}
    \hfill
    \begin{minipage}[t]{0.42\linewidth}
        \centering
        \includegraphics[width=0.9\linewidth]{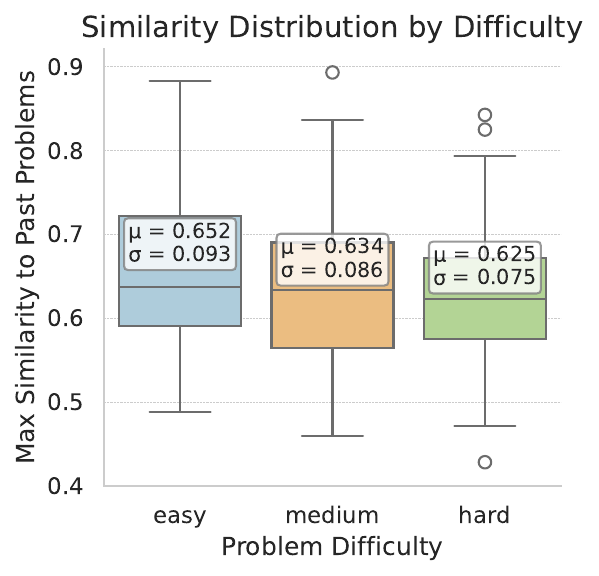}
    \end{minipage}
    \vspace{-15pt}
    \caption{
        \textbf{Left:} Models with higher capacity (High) maintain relatively high performance even on low-similarity problems, while lower-capacity models benefit more from high-similarity problems.
        \textbf{Right:} Easy problems show slightly higher maximum similarity to past problems than hard ones, but the difference is limited.
    }
    \label{fig:reasoning-gap}
    \vspace{-5pt}
\end{figure}

\label{model_vs_similarity}
To examine how model performance relates to problem similarity, we analyze evaluation results from OpenAI’s \texttt{O3-Mini} and \texttt{O4-Mini} variants as reported in LiveCodeBench. Each model is available in three modes—\textit{Low}, \textit{Medium}, and \textit{High}—which correspond to increasing levels of resource consumption ({\it e.g.}, longer context length, higher inference cost, and possibly larger internal activations). As shown in the left panel of Figure~\ref{fig:reasoning-gap}, we observe that the performance gap between these variants narrows as the maximum similarity to past problems increases, and nearly vanishes in the highest bin (0.80–0.90). In contrast, on low-similarity problems, higher-capacity models (especially the \textit{High} variants) maintain strong pass rates, while lower-capacity models degrade significantly. This suggests that high-similarity problems may allow weaker models to perform well by potentially leveraging memorized patterns or surface-level matching, while low-similarity problems are more likely to expose differences in model capability—such as generalization and robustness in reasoning. The right panel shows that easy problems have slightly higher similarity to past data than hard ones, but the difference is modest. This indicates that similarity is not determined by difficulty alone, and should be treated as an independent factor when constructing benchmarks.

\section{Limitations and Future Work}

\noindent
\textbf{Ongoing test set leakage.}
Although we use the most recent problems (from 2023–2024) to construct temporally-separated test sets for code-centric tasks, these examples are likely to appear in future model training corpora as web-scale datasets are updated. As a result, benchmark performance may become inflated over time, even without intentional misuse.
To address this, we plan to release regular updates—every 6 to 12 months—to maintain a clean test split while ensuring the retrieval corpus remains diverse and representative.

\noindent
\textbf{Incomplete and emerging duplicate problems.}
Despite 700 annotated pairs in our Problem-to-Duplicate task, many duplicates—past and future—remain undiscovered, especially on decentralized or low-resource platforms. To address this, we are developing an open-source retrieval platform that supports community-driven annotations and helps contest organizers detect redundancy pre-publication.


\noindent
\textbf{Limited analysis beyond programming.}
While our study focuses on competitive programming, duplication and high similarity may also affect evaluation in other domains, such as mathematical competitions ({\it e.g.}, IMO, AIMO). These issues remain underexplored despite their potential to undermine fair assessment. We encourage future work to examine problem reuse and redundancy in such benchmarks.

\section*{Acknowledgments}

This work was supported by the National Natural Science Foundation of China (Grant No. 62402264), and by the JC STEM Lab of AI for Science and Engineering, funded by The Hong Kong Jockey Club Charities Trust and the Research Grants Council of Hong Kong (Project No. CUHK14213224).

\newpage

\bibliographystyle{unsrt}




\newpage
\appendix

\section{Technical Appendices and Supplementary Material}

\subsection{Additional Experimental Setup Details}
\label{appendix:exp_setup}

For our contrastive pretraining, we use a batch size of 1024, enabled by gradient caching\cite{gradcache} to optimize GPU memory usage. The model is trained with a learning rate of $3 \times 10^{-8}$, temperature $\tau = 0.07$, and weight decay of 0.01, using the AdamW\cite{AdamW} optimizer.  Each query is paired with $m=16$ positive samples, randomly sampled from the dataset. If fewer than $m$ positive samples are available, duplicates are used to pad the group. The maximum input length is set to 1024 tokens. The model is trained for 20 epochs on 8 \texttt{NVIDIA A800} GPUs, taking approximately 70 hours to complete. This stage focuses primarily on problem-code representation alignment. The resulting model, which demonstrates strong performance on code-related retrieval tasks, is denoted as \textbf{CP-Retriever-Code}.

To improve the model's performance on problem-problem retrieval tasks, we perform a second-stage fine-tuning using a multi-task setup. We combine training data from the \textit{Problem-to-Duplicate} and \textit{Simplified-to-Full} tasks, downsampling each to approximately 1,000 examples to ensure balanced representation. The two subsets are then mixed to form the final fine-tuning dataset. This stage uses a batch size of 1, learning rate of $2 \times 10^{-6}$, and weight decay of 0.01. The model is trained for 1 epoch, which takes approximately 1 hour on a single \texttt{NVIDIA A800} GPU. While this model, referred to as \textbf{CP-Retriever-Prob}, slightly underperforms CP-Retriever-Code on code-related tasks, it shows marked improvements in tasks involving problem-level semantic similarity.

For experiments conducted with the \texttt{Qwen3-Embedding-4B} model, we adjusted several hyperparameters to better accommodate its larger architecture. Specifically, the maximum input length was increased to 2048 tokens, the contrastive learning temperature was set to $\tau = 0.05$ (compared to the model’s default of 0.01), and the learning rate was set to $1 \times 10^{-6}$. The total training time scaled approximately proportionally with the increase in model parameters.

\subsubsection{Ablation Study on Group-InfoNCE}
\label{stage1:ablation}

\begin{table}[htbp]
\centering
\caption{\textbf{Performance comparison under different loss functions in contrastive learning.}}
\renewcommand{\arraystretch}{1.2}
\begin{tabular}{@{}lccccc@{}}
\toprule
\textbf{Loss} & \textbf{m (Positives)} & \textbf{Data Aug} & \textbf{Text-to-Code} & \textbf{Code-to-Code} \\
\midrule
Base                     & /   & /     & 39.60 & 68.05 \\
InfoNCE                  & 1   & \xmark & 68.64 & 67.58 \\
InfoNCE                  & 1   & \cmark  & 68.84 & 68.97 \\
InfoNCE (multi-pos)      & 16  & \cmark  & 66.44 & 66.62 \\
Group-InfoNCE            & 4   & \xmark & 69.00 & 69.18 \\
Group-InfoNCE            & 4   & \cmark  & 70.11 & 70.42 \\
Group-InfoNCE            & 16  & \xmark & 69.47 & 69.77 \\
Group-InfoNCE            & 16  & \cmark  & \textbf{70.40} & \textbf{70.59} \\
\bottomrule
\end{tabular}
\label{tab:stage1-ablation}
\end{table}

Table~\ref{tab:stage1-ablation} reports the performance of different loss functions and configurations on the Text-to-Code and Code-to-Code retrieval tasks. 
Here, InfoNCE corresponds to Equation~\ref{eq:infoNCE}, InfoNCE (multi-pos) refers to its multi-positive extension (Equation~\ref{eq:multipos}), and Group-InfoNCE denotes our proposed loss (Equation~\ref{eq:groupinfonce}). 

We observe the following:
\begin{itemize}
    \item Data augmentation consistently improves performance across comparable configurations, highlighting its robustness.
    \item Naïvely increasing positive samples in InfoNCE (multi-pos) does not guarantee gains, suggesting that grouping structure is essential for effective multi-positive learning.
    \item Larger $m$ values in Group-InfoNCE generally lead to higher accuracy, indicating that the model benefits from richer positive context.
\end{itemize}

\begin{table}[htbp]
\centering
\caption{\textbf{Multi-task training validation using Qwen3-4B-Embedding.}}
\renewcommand{\arraystretch}{1.2}
\begin{tabular}{@{}ccccccc@{}}
\toprule
\textbf{m} & \textbf{Problem-Level Finetuning} & \textbf{T2C} & \textbf{C2C} & \textbf{P2Dup} & \textbf{S2Full} & \textbf{Avg} \\
\midrule
No  & No  & 62.68 & 65.15 & 57.23 & 89.31 & 68.59 \\
1   & No  & 67.94 & 66.31 & 56.21 & 90.23 & 70.17 \\
4   & No  & 68.30 & 66.20 & 56.29 & 90.24 & 70.26 \\
16  & No  & 68.54 & 66.49 & 56.43 & 90.07 & 70.38 \\
16  & Yes & 65.85 & 70.18 & 71.58 & 95.02 & \textbf{75.66} \\
\bottomrule
\end{tabular}
\label{tab:qwen4b-multitask}
\end{table}

We further validated the effectiveness of Group-InfoNCE by repeating multi-task training with the \texttt{Qwen3-4B-Embedding} model using the same hyperparameter settings as in our main experiments (Table~\ref{tab:qwen4b-multitask}).
Results show that fine-tuning on problem-level tasks slightly decreases Text-to-Code performance (from 68.54 to 65.85) but substantially improves problem-level tasks such as Problem-to-Duplicate and Simplified-to-Full. 
This consistent, minor trade-off supports our hypothesis that the embedding requirements of code-level and problem-level tasks are inherently different. 
Given this intrinsic conflict, we consider this small drop reasonable and do not pursue extensive hyperparameter optimization that might harm generalizability.

Overall, both ablation and multi-task analyses demonstrate that Group-InfoNCE effectively balances representation learning across heterogeneous tasks while maintaining strong cross-domain retrieval performance.

\subsubsection{Ablation Study on Fine-Tuning Data Composition}

\label{stage2:ablation}

\begin{table*}[htbp]
\centering
\caption{
Ablation results on contrastive pretraining and fine-tuning data composition.  
\textbf{PCD} indicates whether the model was pretrained on CPRet-PCPCD using contrastive learning.  
\textbf{Dup} and \textbf{Simp} indicate the inclusion of Problem-to-Duplicate and Simplified-to-Full data during fine-tuning, respectively.  
The fourth column (\textbf{PCD}) denotes whether CPRet-PCPCD data was also included in the fine-tuning phase.  
T2C = Text-to-Code, C2C = Code-to-Code, P2Dup = Problem-to-Duplicate, S2Full = Simplified-to-Full.
}
\label{tab:ablations}
\begin{tabular}{cccccccccc}
\toprule
\multicolumn{1}{c}{\textbf{T-1}} & \multicolumn{3}{c}{\textbf{T-2}} & \multicolumn{5}{c}{} \\
\cmidrule(lr){1-1} \cmidrule(lr){2-4}
\textbf{PCD} & \textbf{Dup} & \textbf{Simp} & \textbf{PCD} & 
\textbf{T2C} & \textbf{C2C} & \textbf{P2Dup} & \textbf{S2Full} & \textbf{Avg} \\
\midrule
\xmark & \xmark & \xmark & \xmark & 39.60 & 68.05 & 45.26 & 86.43 & 59.84 \\
\xmark & \cmark & \cmark & \xmark & 27.25 & 64.48 & 53.04 & 85.78 & 57.64 \\
\rowcolor{gray!10}
\cmark & \xmark & \xmark & \xmark & 70.40 & 70.59 & 38.68 & 81.45 & 65.28 \\
\rowcolor{gray!10}
\cmark & \cmark & \cmark & \xmark & 56.50 & 70.68 & \textbf{60.06} & 90.74 & 69.50 \\
\cmark & \cmark & \xmark & \xmark & 60.90 & 70.46 & 57.84 & 89.13 & 69.58 \\
\cmark & \xmark & \cmark & \xmark & 68.05 & \textbf{72.77} & 57.01 & \textbf{92.65} & \textbf{72.62} \\
\cmark & \xmark & \cmark & \cmark & \textbf{71.15} & 71.00 & 41.77 & 87.75 & 67.92 \\
\cmark & \cmark & \cmark & \cmark & 70.58 & 70.76 & 50.81 & 88.27 & 70.11 \\
\bottomrule
\end{tabular}
\end{table*}

\noindent
\textbf{Ablation Study Overview.}
Table~\ref{tab:ablations} summarizes a comprehensive ablation study on two key factors in our training pipeline: (I) contrastive pretraining on CPRet-PCPCD, and (II) the composition of fine-tuning data for problem-level tasks.

\noindent
\textbf{Effectiveness of Contrastive Pretraining.}
Rows 3–8 show that models trained with contrastive pretraining significantly outperform those without it (rows 1–2) across all four tasks. Although pretraining alone underperforms on problem-level tasks like \textit{P2Dup} and \textit{S2Full}, models fine-tuned from this initialization consistently surpass those trained from scratch. This indicates that contrastive pretraining helps learn transferable representations. The initial gap on problem-level tasks likely stems from objective mismatch—Group-InfoNCE focuses on problem-code alignment rather than direct problem-to-problem similarity—but still provides a strong foundation for downstream fine-tuning.

\noindent
\textbf{Impact of Fine-Tuning Data Composition.}
Rows 4–8 explore different combinations of \textit{Problem-to-Duplicate}, \textit{Simplified-to-Full}, and optionally CPRet-PCPCD data during fine-tuning. Using only \textit{Simplified-to-Full} (row 6) yields the best overall performance, improving \textit{C2C} and \textit{S2Full}, while maintaining strong results on \textit{P2Dup}.
In contrast, adding \textit{P2Dup} data (row 4) improves that task but slightly reduces generalization elsewhere. Including CPRet-PCPCD during fine-tuning (row 7, 8) helps recover performance on code-centric tasks but slightly weakens results on problem-level tasks.

\noindent
\textbf{Conclusion.}
These results confirm the value of task-specific optimization. We therefore release two final models:
\begin{itemize}
\item \textbf{CPRetriever-Code} (row 3): trained only with contrastive pretraining, optimized for code-centric tasks.
\item \textbf{CPRetriever-Prob} (row 4): further fine-tuned on problem-level tasks, optimized for problem retrieval.
\end{itemize}

\subsection{Supplementary Analysis for Figure 2}

Figure~\ref{fig:temporal-analysis} in the main paper presents two complementary perspectives on model performance trends. The left sub-plot focuses on illustrating the overall improvement of embedding models over time, highlighting representative models that achieved state-of-the-art (SOTA) performance at their respective publication dates. In contrast, the right sub-plot was designed to examine potential data leakage effects by including more recent, high-performing models that are more likely to exhibit such phenomena. As a result, the model sets in the two sub-figures are not identical.

To provide a more comprehensive view, Table~\ref{tab:fig2-supp} below reports the year-wise retrieval performance of additional models (e.g., \texttt{CodeSage-large}, \texttt{Contriever-MSMarco}, and \texttt{E5-Base-v2}) that were not included in the right sub-plot of Figure~\ref{fig:temporal-analysis}. Despite having lower absolute performance compared to the models shown in the right plot, these models display a consistent “decline-then-recovery” trajectory: strong performance in earlier years (e.g., 2015–2016), followed by a sharp degradation (2017–2019), and a partial recovery or stabilization in later years (2020–2024). 

This trend mirrors the temporal performance decay pattern attributed to data leakage, as discussed in Section~\ref{sec:leakage}. Specifically, the observed recovery phase suggests that recently released problems are less contaminated by training data, leading to more realistic evaluations. Therefore, these supplementary results reinforce our main claim that data leakage can substantially inflate retrieval metrics in temporally overlapping benchmarks, and that rigorous temporal separation is critical for fair assessment.

\begin{table}[h]
\centering
\scriptsize
\caption{Supplementary results supporting Figure~\ref{fig:temporal-analysis}. Retrieval performance (\%) of representative models across problem release years. The “decline-then-recovery” pattern indicates potential data leakage in earlier problems.}
\label{tab:fig2-supp}
\begin{tabular}{lcccccccccc}
\toprule
\textbf{Model} & 2015 & 2016 & 2017 & 2018 & 2019 & 2020 & 2021 & 2022 & 2023 & 2024 \\
\midrule
\texttt{SFR-Emb.-Code-2B} & 55.38 & 57.67 & 48.79 & 44.78 & 42.55 & 41.59 & 37.42 & 34.38 & 34.08 & 34.78 \\
\texttt{GTE-ModernBERT-Base} & 42.10 & 45.92 & 28.44 & 15.45 & 10.34 & 12.68 & 9.44 & 9.82 & 9.31 & 9.98 \\
\texttt{CodeSage-Large} & 40.65 & 43.57 & 27.71 & 14.67 & 9.10 & 11.76 & 9.33 & 7.67 & 7.21 & 8.08 \\
\texttt{Contriever-MSMarco} & 16.03 & 19.85 & 10.18 & 3.72 & 1.62 & 1.98 & 0.95 & 0.88 & 0.79 & 0.91 \\
\texttt{E5-Base-v2} & 22.35 & 26.15 & 14.02 & 5.16 & 2.27 & 2.59 & 1.69 & 1.32 & 1.18 & 1.29 \\
\bottomrule
\end{tabular}
\end{table}


\subsection{Evaluation of Retrieval for RAG-based Competitive Programming}
\label{appendix:rag}

To investigate the impact of problem retrieval on downstream code generation tasks, we conducted experiments using Retrieval-Augmented Generation (RAG) in competitive programming problem-solving. We used problems from LiveCodeBench as the test set. To ensure temporal fairness, the retrieval pool was restricted to problems in our dataset published before April 2023, while LiveCodeBench problems appeared starting May 2023. 

For each test problem, we retrieved the top-K most similar problems from the historical subset and provided their descriptions along with one accepted solution as context to the LLM. Three retrieval settings were compared: no retrieval (baseline), using Qwen3-Embedding-4B as the retriever, and using our \textbf{CPRetriever-Prob} (based on Qwen3-Embedding-4B).

For the code generation evaluation, we used three models from the Qwen-Coder series:

\begin{itemize}
    \item \texttt{qwen-coder-turbo} and \texttt{qwen-coder-plus}: non-reasoning models,
    \item \texttt{qwen3-coder-plus}: a reasoning-capable model.
\end{itemize}

The results are summarized in Table~\ref{tab:rag-results}.

\begin{table}[h]
\centering
\caption{\textbf{Pass@1 performance for RAG-based code generation with different retrievers.}}
\begin{tabular}{l l l c}
\toprule
\textbf{Model} & \textbf{Type} & \textbf{RAG Model} & \textbf{Pass@1} \\
\midrule
qwen-coder-turbo & Non-reasoning & None & 0.3575 \\
qwen-coder-turbo & Non-reasoning & Qwen3-Embedding-4B & 0.3825 \\
qwen-coder-turbo & Non-reasoning & CPRetriever-Prob & 0.4075 \\
qwen-coder-plus & Non-reasoning & None & 0.5125 \\
qwen-coder-plus & Non-reasoning & Qwen3-Embedding-4B & 0.5275 \\
qwen-coder-plus & Non-reasoning & CPRetriever-Prob & 0.5500 \\
qwen3-coder-plus & Reasoning & None & 0.7675 \\
qwen3-coder-plus & Reasoning & Qwen3-Embedding-4B & 0.7750 \\
qwen3-coder-plus & Reasoning & CPRetriever-Prob & 0.7800 \\
\bottomrule
\end{tabular}
\label{tab:rag-results}
\end{table}

As shown, incorporating RAG improves Pass@1 across all evaluated models. Non-reasoning models benefit more noticeably, while reasoning models achieve smaller gains due to their strong baseline performance. Notably, using \textbf{CPRetriever-Prob} consistently outperforms Qwen3-Embedding-4B across all LLMs. 

These results demonstrate that effective problem retrieval can enhance competitive programming problem-solving in a RAG setting, highlighting a valuable downstream application for our retrieval model.


\subsection{Challenges in Competitive Programming Problem Retrieval}
\label{appendix:cp-challenges}

Retrieving competitive programming (CP) problems poses distinct challenges compared to standard code retrieval tasks in software engineering (SE) benchmarks such as SWE-Bench. The core difficulty lies in bridging the large semantic gap between a problem's narrative description and its abstract algorithmic solution, whereas SE tasks typically involve more concrete, context-specific code retrieval.

CP problem statements are often highly abstract and indirect. They are framed as puzzles embedded within narratives (e.g., "a farmer needing a route") and require deep semantic understanding to map natural language to algorithmic concepts like graph traversal or dynamic programming. Critical information such as data constraints ($N \le 1{,}000{,}000$) or time limits is subtly embedded, yet it fundamentally dictates the algorithmic complexity required ($O(N^2)$ vs. $O(N \log N)$). Similarly, user queries are abstract (e.g., "longest increasing subsequence sum"), in contrast to SE queries, which are typically specific and grounded in concrete code modifications or bug fixes.

The mapping from a CP problem to its solutions is also more complex. A single problem may admit multiple fundamentally different correct algorithmic approaches (e.g., BFS/DFS vs. dynamic programming for tree diameter). CP retrieval aims to uncover the underlying idea or technique that can inspire a solution strategy, rather than locating an exact code snippet. In SE retrieval, the goal is generally to find specific functions, modules, or API calls that can be directly reused in a known context.

In summary, CP retrieval resembles a mathematician exploring theorems for similar logical structures while ignoring superficial details, focusing on abstract strategies. SE retrieval, in contrast, is more like a mechanic finding the right replacement part for a known machine. This higher-level abstraction and the need to translate between human language and algorithmic logic make competitive programming problem retrieval uniquely challenging.

\subsection{Further Details on the CPRet-PCPCD}
\label{appendix:dataset_stats}

\subsubsection{Language and Problem Format Statistics}

\begin{figure}[h]
    \centering
    \begin{subfigure}{0.45\linewidth}
        \centering
        \includegraphics[width=0.9\linewidth]{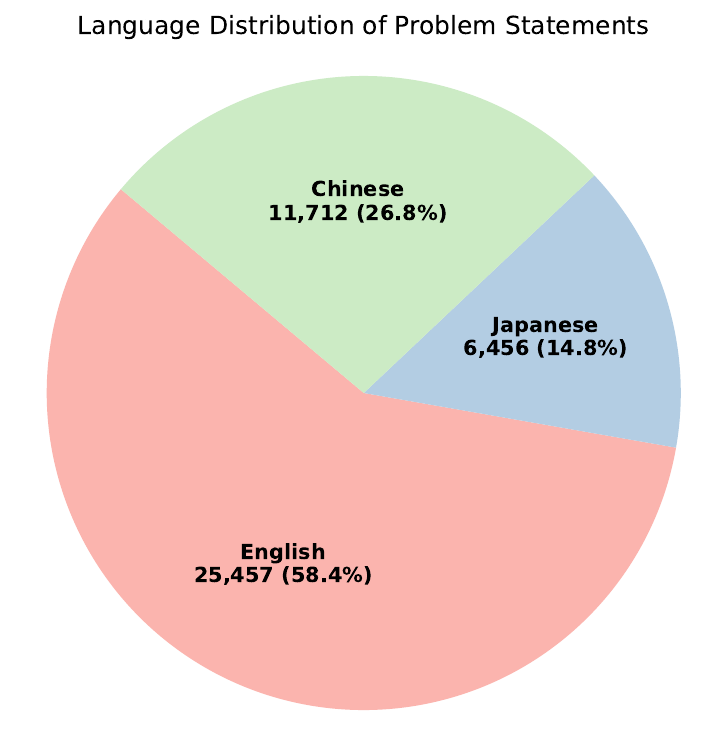}
        \caption{Languages used in problem statements.}
        \label{fig:lang-dist}
    \end{subfigure}
    \hfill
    \begin{subfigure}{0.45\linewidth}
        \centering
        \includegraphics[width=0.9\linewidth]{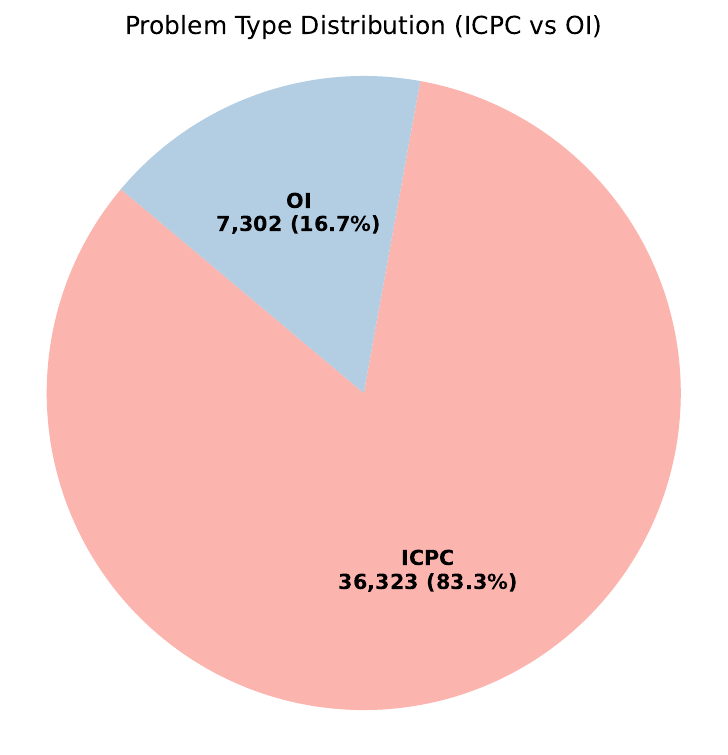}
        \caption{ICPC-style vs OI-style grading.}
        \label{fig:type-dist}
    \end{subfigure}
    \vspace{-0.5em}
    \caption{Dataset-level characteristics by language and format in CPRet-PCPCD.}
    \label{fig:lang-and-type-dist}
    \vspace{-1em}
\end{figure}

Figure~\ref{fig:lang-dist} shows that English is the most prevalent language, largely due to the influence of prior datasets like \textit{TACO}, while Chinese and Japanese examples mainly come from LibreOJ/Nowcoder and AtCoder.  
Figure~\ref{fig:type-dist} reveals that most problems adopt the ICPC-style full-score format, but a notable subset includes OI-style problems that offer partial scores through sub-tasks or test case breakdowns.

\subsubsection{Programming Language Distribution}

\begin{figure}[htbp]
    \centering
    \includegraphics[width=1\linewidth]{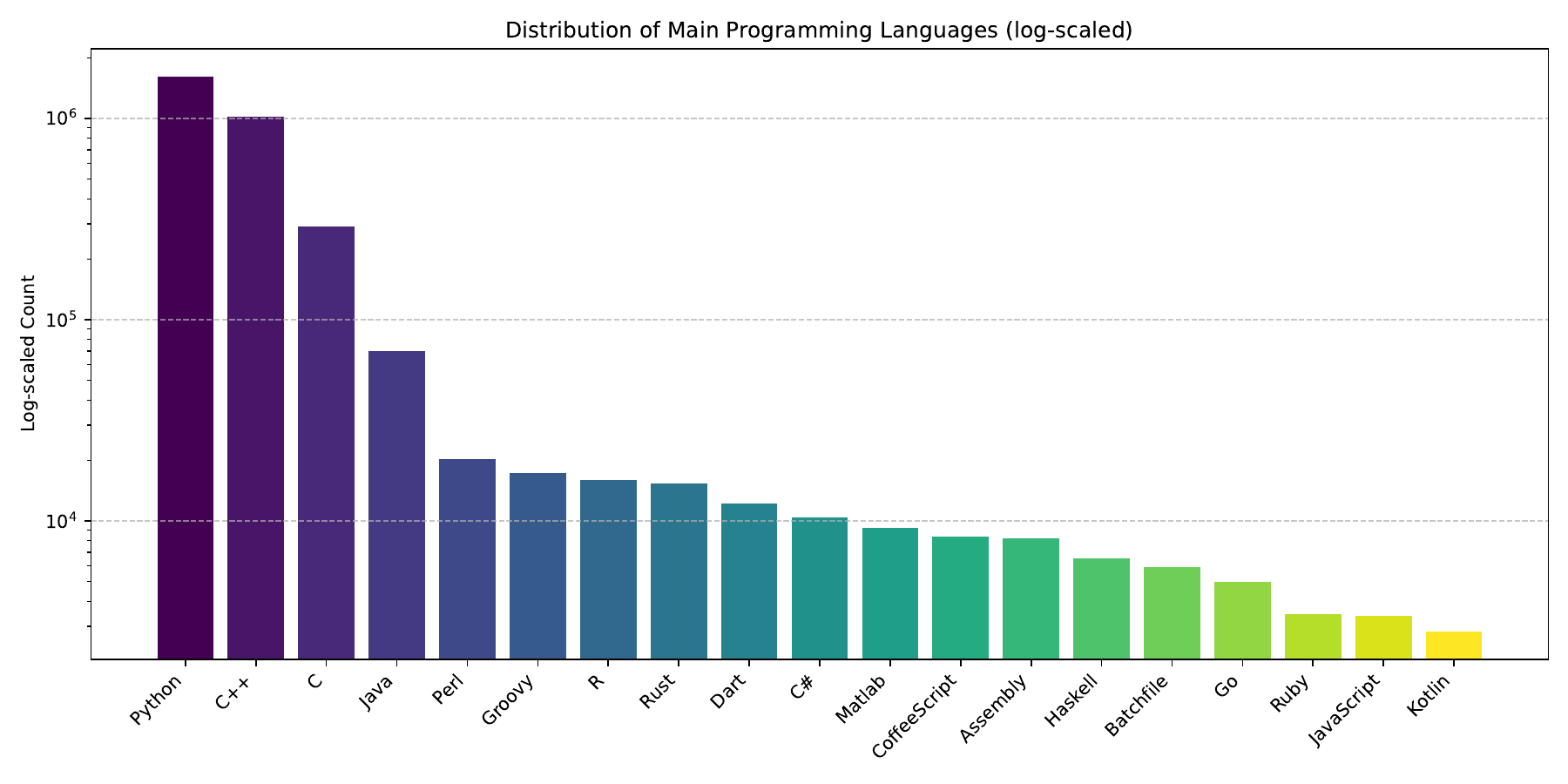}
    \caption{Distribution of programming languages in our dataset CPRet-PCPCD (log-scaled). 
    Python is dominant due to its prevalence in prior datasets like APPS, CodeContests, and TACO. 
    In contrast, our newly collected data is mainly in C/C++, which reflects the practical language choices in competitive programming.}
    \label{fig:language-distribution}
\end{figure}

As shown in Figure~\ref{fig:language-distribution}, our dataset covers a wide range of programming languages.
Early datasets such as \textit{APPS}, \textit{CodeContests}, and \textit{TACO} focused primarily on Python, driven by the popularity of Python in code generation research and the convenience it offers for model training.
However, in our newly collected dataset, the dominant languages shift toward \textbf{C} and \textbf{C++}, which are widely favored in actual competitive programming due to their execution efficiency and low-level control.

Other languages such as \textbf{Java}, \textbf{Perl}, \textbf{Groovy}, and \textbf{Rust} also appear in notable quantities, indicating broader diversity and coverage in our collection.

\subsection{Data Source Details}

\noindent
{\bf Data Copyright Notice.}
All data collected in this work are publicly accessible and do not involve any private or user-specific information. However, the copyright of the original content remains with the respective platforms from which the data were obtained. We use these data solely for research purposes, in accordance with the terms of use and fair academic practice.

\subsubsection{Competitive Programming Problems and Codes Data(CPRet-PCPCD)}

\paragraph{Existing Sources.}
We incorporate the following publicly available datasets:
\begin{itemize}
    \item \textbf{APPS} \cite{apps}: \url{https://github.com/hendrycks/apps}
    \item \textbf{CodeContests} \cite{alphacode}: \url{https://github.com/google-deepmind/code_contests}
    \item \textbf{TACO} \cite{TACO}: \url{https://huggingface.co/datasets/BAAI/TACO}
    \item \textbf{Codeforces Submissions} : \url{https://www.kaggle.com/datasets/yeoyunsianggeremie/codeforces-code-dataset}
    \item \textbf{Codechef Submissions} : \url{https://www.kaggle.com/datasets/arjoonn/codechef-competitive-programming}
    \item \textbf{Codeforces Source Code} : \url{https://www.kaggle.com/datasets/agrigorev/codeforces-code}
\end{itemize}

\noindent
{\bf Newly Collected Sources.}
To enrich the dataset, we additionally collect problems and code from the following platforms:
\begin{itemize}
    \item \textbf{Codeforces}(English, ICPC): \url{https://codeforces.com/}, with API reference at \url{https://codeforces.com/apiHelp}
    \item \textbf{AtCoder}(English and Japanese, ICPC + OI): \url{https://atcoder.jp/}, with API reference at \url{https://github.com/kenkoooo/AtCoderProblems/blob/master/doc/api.md}
    \item \textbf{LibreOJ}(Chinese, ICPC + OI): \url{https://loj.ac/}, with API reference at \url{https://api.loj.ac/}
    \item \textbf{Nowcoder}(Chinese, ICPC + OI): \url{https://ac.nowcoder.com/}
\end{itemize}

For these new sources, we typically first obtain problem and submission metadata via the platform’s API (when available), and then download actual code submissions accordingly. In cases where APIs are not provided, we resort to web crawling.

\noindent
{\bf Crawler Tools.}
We utilize the following libraries to implement our crawlers:
\begin{itemize}
    \item \texttt{curl\_cffi}: \url{https://github.com/lexiforest/curl_cffi}
    \item \texttt{crawl4ai}: \url{https://github.com/unclecode/crawl4ai}
\end{itemize}

To minimize load on the original platforms and avoid abuse, we do not publicly release our custom crawlers. For academic collaboration or access, please contact us via email.

\subsubsection{Duplicate Programming Problem Pairs}

To collect semantically similar or duplicate programming problem pairs, we crawled discussion forum data from the following two competitive programming platforms:

\begin{itemize}
    \item \textbf{Codeforces}, via blog entries such as \url{https://codeforces.com/blog/entry/142637}
    \item \textbf{Luogu}, via forum posts such as \url{https://www.luogu.com.cn/discuss?forum=P1000}
\end{itemize}

From Codeforces, we gathered approximately 130,000 blog posts, each containing an average of 12 user comments.  
From Luogu, we collected around 236,000 discussion comments.

To extract high-quality duplicate problem pairs, we use a multi-stage filtering process combining:
\begin{itemize}
    \item \textbf{Keyword filtering} to identify relevant discussions,
    \item \textbf{Large Language Model (LLM)-based scoring} for candidate selection, and
    \item \textbf{Manual annotation} to ensure correctness and semantic equivalence.
\end{itemize}

\noindent
{\bf Note.}
For each problem pair mentioned in the forums, both problems may originate from the same platform, or from different platforms. In most cases, at least one problem in the pair comes from the platform where the discussion was found.

\subsubsection{Simplified and Full Problem Description Pairs}

This dataset is collected entirely from the Luogu platform (https://www.luogu.com.cn/), where users have contributed simplified versions of competitive programming problems originally published on international online judges such as Codeforces (https://codeforces.com/), AtCoder (https://atcoder.jp/), SPOJ (https://www.spoj.com/), and UVA (https://onlinejudge.org/). These problems may have been originally written in English, Japanese, or other languages. The simplified versions are often written in Chinese and aim not only to translate the content, but also to restructure and clarify problem statements by removing less essential narrative elements, rephrasing complex sentences, and highlighting key constraints and requirements. This simplification process helps make the problems more accessible to beginners.

We crawled these user-generated simplifications and applied filtering steps to remove low-quality entries, such as those that are incomplete, inconsistent with the original, or overly literal machine translations. To ensure consistency across the dataset, we use \textit{Qwen-2.5-Max}\cite{qwen2} to translate Chinese simplified statements into English before inclusion. The resulting dataset maintains a high level of clarity and fidelity, and serves as a valuable resource for studying cross-lingual and cross-abstraction retrieval.

\subsection{Extended Results: Model Performance across Similarity Bins}

\begin{figure}[htbp]
    \centering
    \includegraphics[width=1.0\linewidth]{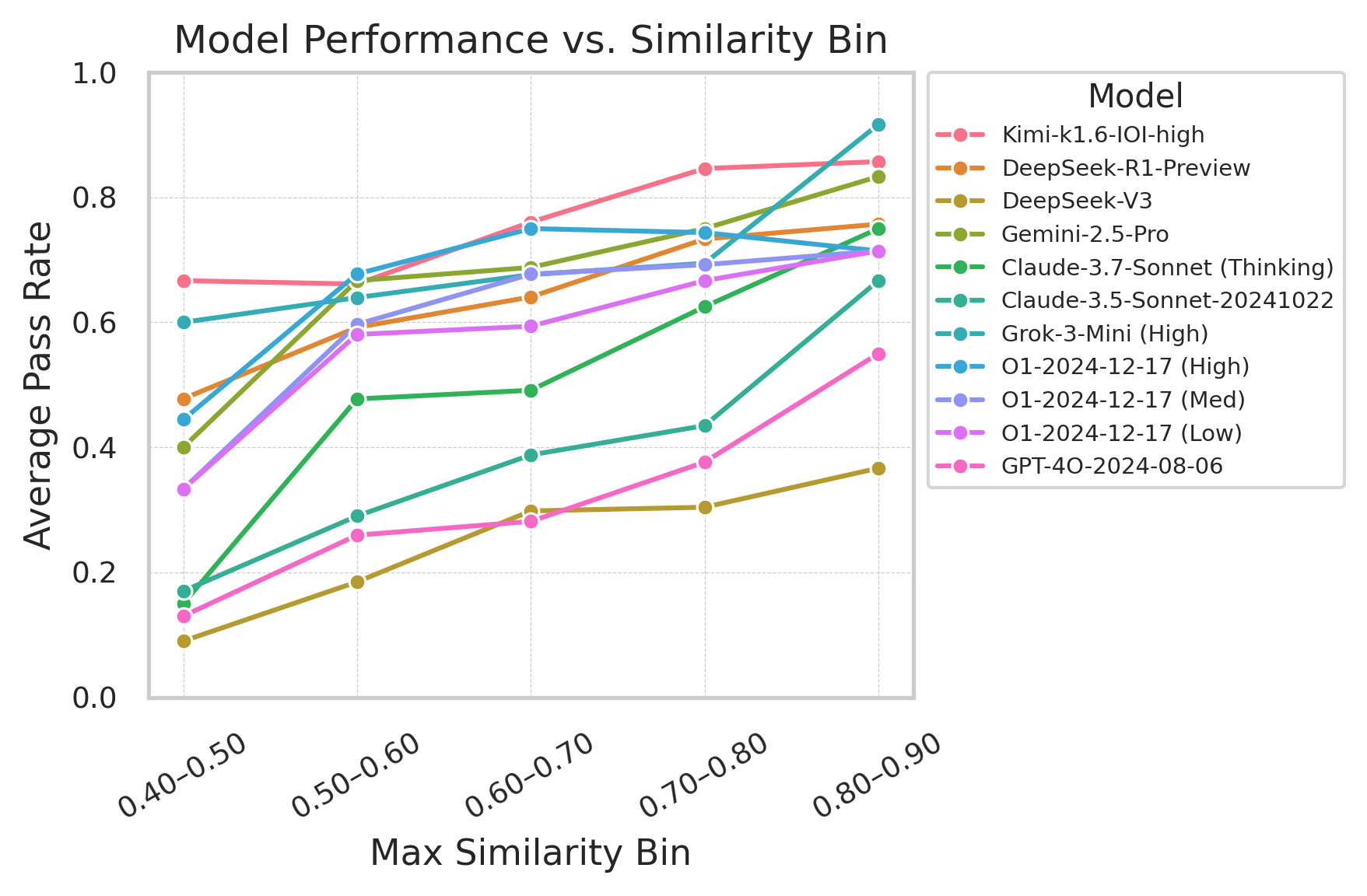}
    \caption{
        \textbf{Extended results on model performance vs. similarity.}
    }
    \label{fig:appendix-sim-bin}
\end{figure}

Figure~\ref{fig:appendix-sim-bin} expands on the similarity-based analysis in Section~\ref{model_vs_similarity}, showing average pass rates across max similarity bins for additional models not included in the main paper.

The observed trends further support the conclusion that higher retrieval similarity strongly correlates with code generation success, demonstrating consistency across different model scales and types.

\subsection{Cases of Retrieval Tasks}

\subsubsection{Text-to-Code and Code-to-Code}

We next present a problem along with two correct solutions.
In the Text-to-Code task, the goal is to retrieve these solutions given the problem description.
In the Code-to-Code task, the goal is to retrieve the other solution given one of them, which may differ in language or implementation.

\begin{tcolorbox}[colback=blue!1!white, colframe=blue!30!black, boxrule=0.5pt, arc=2pt,
    left=4pt, right=4pt, top=4pt, bottom=4pt, fontupper=\ttfamily\small,
    title=https://atcoder.jp/contests/arc162/tasks/arc162\_a]

\textbf{Problem Statement}

There are $N$ people, numbered from $1$ to $N$, who participated in a round-trip race between two points. The following information is recorded about this race.

\begin{itemize}
    \item The \textbf{outward} times of any two people were different, and person $i$ $(1 \le i \le N)$ had the $i$-th fastest outward time.
    \item The \textbf{round-trip} times (the sum of the outward and return times) of any two people were different, and person $i$ $(1 \le i \le N)$ had the $P_i$-th fastest round-trip time.
    \item The person (or persons) with the fastest \textbf{return} time was awarded the \textbf{fastest return award}.
\end{itemize}

Here, $P_1, P_2, \dots, P_N$ is a permutation of $1, 2, \dots, N$.

How many people could have received the \textbf{fastest return award}?

There are $T$ test cases. Answer each of them.

\vspace{0.5em}
\textbf{Constraints}
\begin{itemize}
    \item $1 \le T \le 500$
    \item $2 \le N \le 10^3$
    \item $P_1, P_2, \dots, P_N$ is a permutation of $1, 2, \dots, N$
    \item All input values are integers
    \item The sum of $N$ over all test cases does not exceed $10^3$
\end{itemize}

\vspace{0.5em}
\textbf{Input Format}

The input is given from standard input in the following format:
\begin{verbatim}
T
case_1
...
case_T
\end{verbatim}

Each test case $i$ $(1 \le i \le T)$ is given in the following format:
\begin{verbatim}
N
P_1 P_2 ... P_N
\end{verbatim}

\vspace{0.5em}
\textbf{Output Format}

Print $T$ lines. The $i$-th line $(1 \le i \le T)$ should contain the answer for the $i$-th test case.

\end{tcolorbox}

\lstset{
  basicstyle=\ttfamily\small,
  keywordstyle=\color{blue},
  commentstyle=\color{gray},
  stringstyle=\color{red},
  breaklines=true,
  frame=single
}

\begin{lstlisting}[language=c++, caption={Solution Code 1(C++)}, label={lst:cf-solution}]
#include <bits/stdc++.h>
#include <cstdlib>
#include <math.h>

using namespace std;

int main()
{
    int t;
    cin >> t;
    while(t--){
        int n;
        cin >> n;
        vector<int>a(n);
        vector<int>b(n);
        for(int i=0;i<n;i++){
            cin >> a[i];
            a[i]--;
            b[a[i]]=i;
        }
        int ans=0;
        for(int i=0;i<n;i++){
            bool ok=true;
            for(int j=0;j<i;j++){
                if(b[i]<b[j])ok=false;
            }
            if(ok)ans++;
        }
        cout << ans << endl;
    }
    return 0;
}
\end{lstlisting}

\begin{lstlisting}[language=Ruby, caption={Solution Code 2(Ruby)}]
t = read_line.to_i

def solve
  n = read_line.to_i
  p = read_line.split.map { |x| x.to_i - 1 }

  n.times.count do |i|
    n.times.all? do |j|
      # i < j => p[i] < p[j]
      !(i < j) || p[i] < p[j]
    end
  end
end

t.times do 
  puts solve
end
\end{lstlisting}

\subsubsection{Problem-to-Duplicate}

The only difference between the two problems is that one includes multiple test cases, while the other does not. Aside from this, the problems are nearly identical, and are therefore classified as a Near Match.

\begin{tcolorbox}[colback=blue!1!white, colframe=blue!30!black, boxrule=0.5pt, arc=2pt,
        left=4pt, right=4pt, top=4pt, bottom=4pt, fontupper=\ttfamily\small,
        title=https://codeforces.com/contest/1702/problem/G2]

Problem: G2. Passable Paths (hard version)

You are given a tree with $n$ vertices numbered from $1$ to $n$. A tree is an undirected connected graph with $n - 1$ edges and no cycles.

You are also given $q$ queries. In each query, you are given a set of $k$ vertices. For each query, determine whether there exists a simple path in the tree that passes through all the given $k$ vertices.

\textbf{Input}

The first line contains an integer $n$ ($2 \leq n \leq 2 \times 10^5$) — the number of vertices in the tree.

Each of the next $n - 1$ lines contains two integers $u$ and $v$ ($1 \leq u, v \leq n$) — the endpoints of an edge in the tree.

The next line contains an integer $q$ ($1 \leq q \leq 2 \times 10^5$) — the number of queries.

Each of the next $q$ lines describes a query:
\begin{itemize}
  \item The first integer $k$ ($2 \leq k \leq n$) — the number of vertices in the query.
  \item Followed by $k$ integers $v_1, v_2, \ldots, v_k$ ($1 \leq v_i \leq n$) — the vertices in the query.
\end{itemize}

\textbf{Output}

For each query, print \texttt{YES} if there exists a simple path that passes through all the given $k$ vertices. Otherwise, print \texttt{NO}.
\end{tcolorbox}

\begin{tcolorbox}[colback=blue!1!white, colframe=blue!30!black, boxrule=0.5pt, arc=2pt,
        left=4pt, right=4pt, top=4pt, bottom=4pt, fontupper=\ttfamily\small,
        title=https://www.codechef.com/JULY21A/problems/KPATHQRY]
Problem: KPATHQRY — Path Queries on Trees

You're given a tree with $N$ vertices numbered from $1$ to $N$. Your goal is to handle $Q$ queries. For each query, you are given $K$ nodes $v_1, v_2, \ldots, v_K$. Find whether there exists a simple path in the tree that covers all the given vertices.

\textbf{Input}

\begin{itemize}
    \item The first line contains a single integer $T$ — the number of test cases.
    \item For each test case:
    \begin{itemize}
        \item The first line contains a single integer $N$ — the number of vertices.
        \item Each of the following $N-1$ lines contains two integers $u$ and $v$ — an edge in the tree.
        \item The next line contains a single integer $Q$ — the number of queries.
        \item Each of the following $Q$ lines describes a query:
        \begin{itemize}
            \item Starts with $K_i$ — the number of vertices in the query.
            \item Followed by $K_i$ integers: $v_1, v_2, \ldots, v_{K_i}$.
        \end{itemize}
    \end{itemize}
\end{itemize}

\textbf{Output}

For each query, print \texttt{"YES"} if a simple path covering all the given nodes exists, otherwise print \texttt{"NO"}.

You may print each character of the string in uppercase or lowercase (for example, \texttt{"yEs"}, \texttt{"yes"}, \texttt{"Yes"}, and \texttt{"YES"} will all be treated as identical).

\textbf{Constraints}

\begin{itemize}
    \item $1 \leq T \leq 10$
    \item $1 \leq N \leq 10^5$
    \item $1 \leq u, v, v_j \leq N$
    \item $1 \leq Q \leq 10^5$
    \item $1 \leq K_i \leq N$
    \item The edges form a valid tree.
    \item All vertices in a single query are distinct.
    \item The sum of $N$ over all test cases does not exceed $2 \cdot 10^5$.
    \item The sum of $K_i$ over all queries in a single test case does not exceed $10^5$.
\end{itemize}

\textbf{Subtasks}

\begin{itemize}
    \item \textbf{Subtask \#1 (100 points):} Original constraints.
\end{itemize}

\end{tcolorbox}

\subsubsection{Simplified-to-Full}
The first paragraph below is a simplified version of the second paragraph. The simplification removes irrelevant background information and input/output format details, preserving only the core problem statement.

\begin{tcolorbox}[colback=blue!1!white, colframe=blue!30!black, boxrule=0.5pt, arc=2pt,
        left=4pt, right=4pt, top=4pt, bottom=4pt, fontupper=\ttfamily\small,
        title=https://codeforces.com/problemset/problem/1183/F Concise problem description]
You have $n$ numbers from $a_1$ to $a_n$, and you want to select \textbf{at most} 3 numbers such that no number is a multiple of another. You aim to maximize the sum of the selected numbers. Output this maximum sum.
\end{tcolorbox}

\begin{tcolorbox}[colback=blue!1!white, colframe=blue!30!black, boxrule=0.5pt, arc=2pt,
        left=4pt, right=4pt, top=4pt, bottom=4pt, fontupper=\ttfamily\small,
        title=https://codeforces.com/problemset/problem/1183/F Full problem description]
        
Topforces Strikes Back

Problem Description
\begin{itemize}
    \item One important contest will take place on the most famous programming platform (Topforces) very soon!
    \item The authors have a pool of $ n $ problems and should choose at most three of them into this contest. The prettiness of the $ i $ -th problem is $ a_i $ . The authors have to compose the most pretty contest (in other words, the cumulative prettinesses of chosen problems should be maximum possible).
    \item But there is one important thing in the contest preparation: because of some superstitions of authors, the prettinesses of problems cannot divide each other. In other words, if the prettinesses of chosen problems are $ x, y, z $ , then $ x $ should be divisible by neither $ y $ , nor $ z $ , $ y $ should be divisible by neither $ x $ , nor $ z $ and $ z $ should be divisible by neither $ x $ , nor $ y $ . If the prettinesses of chosen problems are $ x $ and $ y $ then neither $ x $ should be divisible by $ y $ nor $ y $ should be divisible by $ x $ . Any contest composed from one problem is considered good.
    \item Your task is to find out the maximum possible total prettiness of the contest composed of at most three problems from the given pool.
    \item You have to answer $ q $ independent queries.
    \item If you are Python programmer, consider using PyPy instead of Python when you submit your code.
\end{itemize}

Input Format
\begin{itemize}
    \item The first line of the input contains one integer $ q $ ( $ 1 \le q \le 2 \cdot 10^5 $ ) — the number of queries.
    \item The first line of the query contains one integer $ n $ ( $ 1 \le n \le 2 \cdot 10^5 $ ) — the number of problems.
    \item The second line of the query contains $ n $ integers $ a_1, a_2, \dots, a_n $ ( $ 2 \le a_i \le 2 \cdot 10^5 $ ), where $ a_i $ is the prettiness of the $ i $ -th problem.
    \item It is guaranteed that the sum of $ n $ over all queries does not exceed $ 2 \cdot 10^5 $ .
\end{itemize}

Output Format
\begin{itemize}
    \item For each query print one integer — the maximum possible cumulative prettiness of the contest composed of at most three problems from the given pool of problems in the query.
\end{itemize}

\end{tcolorbox}

\subsection{Online Demo and Test Cases of CPRetriever}
\label{appendix:demo}

We deployed an open-source competitive programming problem retrieval platform, \textbf{CPRetriever}, on May 21, 2025 (\url{https://cpret.online/}). Within the first week, the platform processed nearly 2,000 search queries, and the announcement post on Codeforces received over 250 upvotes, indicating strong community interest.

The platform supports two primary retrieval functionalities: (i) \textbf{similar problem retrieval}, which assists users in expanding problem-solving perspectives and identifying knowledge gaps, and (ii) \textbf{duplicate problem detection}, which aids problem setters in identifying previously seen ideas or solutions.

\subsubsection{Duplicate Problem Retrieval in a Recent Contest}

We evaluated CPRetriever on the 2025 CCPC National Invitational Contest (Northeast), which featured 12 problems (\url{https://codeforces.com/gym/105924}). The system successfully identified six problems with highly similar or identical historical counterparts. Manual inspection of the top three retrievals per query suggests a \textbf{minimum duplicate rate of 50\%}, highlighting potential fairness concerns in contest scoring.

\begin{table}[htbp]
\centering
\small
\caption{Detected duplicates in the 2025 CCPC Northeast Contest.}
\begin{tabular}{@{}l l l l@{}}
\toprule
\textbf{Contest Problem} & \textbf{Matched Historical Problem} & \textbf{Similarity Level} & \textbf{Rank} \\
\midrule
A. GD Ultimate Rhythm Lab & Nowcoder - Xiao Rui Rui's Sequence & Same approach & 1 \\
D. Defend the Carrot & SPOJ - UOFTBB & Almost identical & 1 \\
E. Tree Edge Removal & Luogu - [JRKSJ R7] Stem & Almost identical & 1 \\
F. Youthful Oath II & Codeforces - 80B Depression & Almost identical & 1 \\
J. Kingdom: Memories & AtCoder - R Walk & Almost identical & 3 \\
L. Bathhouse & Codeforces - 219E Parking Lot & Same approach & 2 \\
\bottomrule
\end{tabular}
\end{table}

\subsubsection{Similar Problem Retrieval: MEX Variants}

We further evaluated the system using the classic ``interval MEX'' problem to retrieve its variants across multiple contests, demonstrating CPRetriever's utility for \textbf{idea exploration and knowledge transfer}.

\begin{table}[htbp]
\centering
\footnotesize
\caption{Top retrieval results for an interval MEX query.}
\begin{tabular}{@{}l l l@{}}
\toprule
\textbf{Rank} & \textbf{Problem} & \textbf{Description} \\
\midrule
1 & Luogu P4137: RMQ Problem / MEX & Original problem \\
2 & LOJ 6908: THUPC 2024 Prelim - ``Matryoshka'' & MEX of all subarrays of length $k$, then take MEX \\
5 & AtCoder ABC194E: Mex Min & MEX of all subarrays of length $k$, then take minimum \\
6 & Luogu P10032: Mex of Sequence & Repeated operations: $a'[i] = \mathrm{mex}(a \setminus a[i])$ \\
11 & Nowcoder 237670: Classic Problem & MEX queries on permutations, optimized to $O(n + m)$ \\
14 & Luogu P8087: JROI-5 Interval & MEX of all subarrays of length $k$, then take maximum \\
15 & AtCoder ABC290C: Max MEX & MEX of all subsequences of length $k$, then take minimum \\
16 & Codeforces 1436E: Complicated Computations & MEX of all subarrays, then take MEX again \\
23 & AtCoder ABC330E: Mex and Update & Supports element modification or full-array MEX queries \\
24 & Luogu P11837: Making Mexes B & Minimum edits to ensure $\mathrm{mex}(a) = i$ \\
\bottomrule
\end{tabular}
\end{table}

These case studies demonstrate that CPRetriever effectively identifies both \textbf{duplicates} and \textbf{semantically similar problems}, supporting practical contest preparation, knowledge expansion, and fairness monitoring in competitive programming environments.

\newpage
\subsection{Figures of data collection and processing pipeline}

\begin{figure}[h]
\centering
\resizebox{0.2\textwidth}{!}{
\begin{tikzpicture}[node distance=2.2cm]

\node (source) [data] {Problem Sources \\ (Codeforces, AtCoder, Luogu, etc.)};
\node (collect) [process, below of=source] {Crawling and Data Extraction};
\node (pairing) [process, below of=collect] {Match Problems with \\ Accepted Solutions};
\node (meta) [process, below of=pairing] {Collect Metadata: \\ Language, Contest Type, Timestamp};
\node (diversity) [data, below of=meta] {Multi-language \& Multi-format Coverage \\ (ICPC-style, OI-style)};
\node (clean) [process, below of=diversity] {Quality Filtering \\ \& Data Normalization};
\node (annotate) [process, below of=clean] {Annotate with Temporal Information};
\node (output) [data, below of=annotate] {\textbf{CPRet-PCPCD Dataset} \\ (Problems + Solutions + Metadata)};

\draw [arrow] (source) -- (collect);
\draw [arrow] (collect) -- (pairing);
\draw [arrow] (pairing) -- (meta);
\draw [arrow] (meta) -- (diversity);
\draw [arrow] (diversity) -- (clean);
\draw [arrow] (clean) -- (annotate);
\draw [arrow] (annotate) -- (output);

\end{tikzpicture}
}
\caption{Overall construction pipeline of the \textbf{CPRet-PCPCD} dataset. 
Problems and accepted solutions are collected from multiple online judges, paired and enriched with metadata, filtered for quality, and annotated with timestamps to support temporally-aware retrieval research.}
\label{fig:dataset_pipeline}
\end{figure}

\begin{figure}[h]
\centering
\resizebox{0.9\textwidth}{!}{
\begin{tikzpicture}[node distance=2.2cm]

\node (source) [data] {Discussion Threads \& Blog Posts \\ (Codeforces, Luogu)};
\node (heuristic) [process, below of=source] {Keyword-based Filtering \\ \& LLM Classification};
\node (candidates) [data, below of=heuristic] {5K Candidate Entries};
\node (annotation) [process, below of=candidates] {Manual Annotation by \\ Experienced Programmers};
\node (criteria) [data, below of=annotation] {Three Duplication Levels: \\ Exact / Near / Method Match};
\node (cluster) [process, below of=criteria] {Clustering Duplicate Problems};
\node (clusterset) [data, below of=cluster] {700 Duplicate Pairs \\ (Clusters Formed)};
\node (split) [decision, below of=clusterset, yshift=-0.3cm] {Split?};
\node (testset) [data, below left=1.2cm and 2.5cm of split] {30\% Clusters \\ $\rightarrow$ Test Set};
\node (query) [process, below right=1.2cm and 2.5cm of split] {Randomly Select \\ Query \& Corpus Problems};
\node (distractors) [process, below=1.5cm of split] {Add Remaining Codeforces \\ Problems as Distractors};
\node (task) [process, below=1.7cm of distractors] {\textbf{Problem-to-Duplicate} Retrieval Task};

\draw [arrow] (source) -- (heuristic);
\draw [arrow] (heuristic) -- (candidates);
\draw [arrow] (candidates) -- (annotation);
\draw [arrow] (annotation) -- (criteria);
\draw [arrow] (criteria) -- (cluster);
\draw [arrow] (cluster) -- (clusterset);
\draw [arrow] (clusterset) -- (split);
\draw [arrow] (split) -- node[anchor=east] {Select} (testset);
\draw [arrow] (split) -- node[anchor=west] {Select} (query);
\draw [arrow] (testset) |- (distractors);
\draw [arrow] (query) |- (distractors);
\draw [arrow] (distractors) -- (task);

\end{tikzpicture}
}
\caption{Data collection and processing pipeline for the \textbf{Problem-to-Duplicate} retrieval task. 
Discussion threads and blogs are filtered and annotated to identify duplicate problems, which are clustered and split into test sets with added distractors for realistic retrieval evaluation.}
\label{fig:duplicate_pipeline}
\end{figure}

\begin{figure}[h]
\centering
\resizebox{0.9\textwidth}{!}{
\begin{tikzpicture}[node distance=2.1cm]

\node (source) [data] {Original Problems \\ (Codeforces / AtCoder)};
\node (usergen) [process, below of=source] {User-contributed Simplified \\ \& Translated Versions (Luogu)};
\node (crawl) [process, below of=usergen] {Data Crawling};
\node (filter) [process, below of=crawl] {Filtering \& Quality Control \\ (remove low-quality or MT-based entries)};
\node (cleaned) [data, below of=filter] {17K High-quality \\ Simplified--Full Pairs};
\node (split) [decision, below of=cleaned, yshift=-0.3cm] {Split?};
\node (train) [data, below left=1.2cm and 2.5cm of split] {Train Set \\ (7K pairs)};
\node (test) [data, below right=1.2cm and 2.5cm of split] {Test Set \\ (10K pairs)};
\node (task) [process, below=1.6cm of split, xshift=0cm] {\textbf{Simplified-to-Full} Retrieval Task \\ (Query = simplified, Corpus = full)};

\draw [arrow] (source) -- (usergen);
\draw [arrow] (usergen) -- (crawl);
\draw [arrow] (crawl) -- (filter);
\draw [arrow] (filter) -- (cleaned);
\draw [arrow] (cleaned) -- (split);
\draw [arrow] (split) -- node[anchor=east] {70\%} (train);
\draw [arrow] (split) -- node[anchor=west] {30\%} (test);
\draw [arrow] (train) |- (task);
\draw [arrow] (test) |- (task);

\end{tikzpicture}
}
\caption{Data collection and processing pipeline for the \textbf{Simplified-to-Full} retrieval task. 
Problems are simplified or translated by Luogu users, then crawled, filtered for quality, and split into training and test sets.}
\label{fig:simplified2full_pipeline}
\end{figure}

\end{document}